\documentclass[aoas,preprint]{imsart}

\usepackage{graphicx} % support the \includegraphics command and options
\usepackage{epstopdf}
\usepackage{caption}
\usepackage{subcaption}
\usepackage{amsthm,amsmath,natbib}
\usepackage{amssymb}
\RequirePackage[colorlinks,citecolor=blue,urlcolor=blue]{hyperref}
\newcommand{\M}[1]{\boldsymbol{#1}}  %matrix
\newcommand{\V}[1]{\boldsymbol{#1}}  %vector
\arxiv{arXiv:1611.08280}
\usepackage{algorithm,algcompatible,amsmath}
\algnewcommand\INPUT{\item[\textbf{Input:}]}%
\algnewcommand\OUTPUT{\item[\textbf{Output:}]}%

\startlocaldefs
\endlocaldefs
\usepackage[utf8]{inputenc}
\usepackage[english]{babel} 
\newtheorem{theorem}{Theorem}[section]

\begin{document}

\begin{frontmatter}

% "Title of the paper"
\title{Two-Level Structural Sparsity Regularization for Identifying Lattices and Defects in Noisy Images}
%\runtitle{A Sample Document}
%\thankstext{T1}{Footnote to the title with the ``thankstext'' command.}

\begin{aug}
	\author{\fnms{Xin} \snm{Li}\thanksref{m1}\ead[label=e1]{xl12d@my.fsu.edu}},
	\author{\fnms{Alex} \snm{Belianinov}\thanksref{m2}\ead[label=e3]{belianinova@ornl.gov}},
	\author{\fnms{Ondrej} \snm{Dyck}\thanksref{m2}\ead[label=e4]{dyckoe@ornl.gov}},
	\author{\fnms{Stephen} \snm{Jesse}\thanksref{m2}\ead[label=e5]{sjesse@ornl.gov}},
	\and
	\author{\fnms{Chiwoo}  \snm{Park}\thanksref{m1}\ead[label=e2]{cpark5@fsu.edu}}

%	\thankstext{t1}{Some comment}
%	\thankstext{t2}{First supporter of the project}
%	\thankstext{t3}{Second supporter of the project}
	\runauthor{Li, X. et al.}
	
	\affiliation{Florida State University\thanksmark{m1}\\
		 Oak Ridge National Laboratory \thanksmark{m2}}
	
	\address{Address of the First and Last Authors\\
		Department of Industrial and Manufacturing Engineering\\
		Florida State University \\
		2525 Pottsdamer St.\\
		Tallahassee FL 32310-6046\\
		\printead{e1}\\
		\phantom{E-mail:\ }\printead*{e2}}
		
		\address{Address of the Second, Third and Fourth Authors\\
		Center for Nanophase Material Science\\
     	Oak Ridge National Laboratory\\
     	1 Bethel Valley Rd \\
     	Oak Ridge, TN 37830\\     	
		\printead{e3}\\
		\phantom{E-mail:\ }\printead*{e4}\\ 
		\phantom{E-mail:\ }\printead*{e5}}

\end{aug}

%\runauthor{L. Xin et al.}

\begin{abstract}
: This paper presents a regularized regression model with a two-level structural sparsity penalty applied to locate individual atoms in a noisy scanning transmission electron microscopy image (STEM). In crystals, the locations of atoms is symmetric, condensed into a few lattice groups. Therefore, by identifying the underlying lattice in a given image, individual atoms can be accurately located. We propose to formulate the identification of the lattice groups as a sparse group selection problem.  Furthermore, real atomic scale images contain defects and vacancies, so atomic identification based solely on a lattice group may result in false positives and false negatives. To minimize error, model includes an individual sparsity regularization in addition to the group sparsity for a within-group selection, which results in a regression model with a two-level sparsity regularization. We propose a modification of the group orthogonal matching pursuit (gOMP) algorithm with a thresholding step to solve the atom finding problem. The convergence and statistical analyses of the proposed algorithm are presented. The proposed algorithm is also evaluated through numerical experiments with simulated images. The applicability of the algorithm on determination of atom structures and identification of imaging distortions and atomic defects was demonstrated using three real STEM images. We believe this is an important step toward automatic phase identification and assignment with the advent of genomic databases for materials.
\end{abstract}

\begin{keyword} %[class=MSC]
\kwd{sparse regression}
\kwd{structural sparsity}
\kwd{lattice group}
\kwd{structural evaluation of materials}
\kwd{image data analysis}
\end{keyword}

%\begin{keyword}
%\kwd{}
%\kwd{}
%\end{keyword}

\end{frontmatter}

\section{Introduction}
Crystallographic studies at atomic length scales with locating individual atoms, identifying symmetries, dislocations and defects, which impact material properties. Quantification of phase, defect, and interface information allows researchers to contextualize material performance as a function of measurable parameters such as distance, or energy, which can be extracted directly from microscopy data. The ultimate goal of localized imaging and spectroscopy is to observe and quantitatively correlate structure-property relationships with functionality, by evaluating chemical, electronic, optical and phonon properties of individual atomic and nanometer-sized structural elements \citep{mody2011instrumental}. Historic improvements in the underlying instrument hardware and data processing technologies has allowed determination of atomic positions with sub-10 pm precision \citep{yankovich2014picometre,kim2012probing}, which enabled the visualization of chemical and mechanical strains \citep{kim2014direct}, and order parameter fields including ferroelectric polarization \citep{chang2011atomically,nelson2011spontaneous,jia2007unit,jia2011direct} and octahedral tilts \citep{jia2009oxygen,kim2013interplay,borisevich2010suppression,he2010control,borisevich2010mapping}. However, quantifying structural information directly from  images has been a challenge due to a large number of atoms and imaging artifacts. In this work we describe a statistical approach to process these data in order to extract the material structure from atomically resolved images.

The material used as a motivating example throughout this paper is Mo-V-M-O (M = Ta, Nb, Sb, Te) M1 and M2 mixed phase oxides. These complex oxides have recently drawn considerable research attention \citep{shiju2009recent} as they are the most promising catalysts for propane ammoxidation reactions to make acrylonitrile (ACN), an industrially important chemical currently produced on a scale of 6 million tons annually \citep{bradzil2010acrylonitrile}. The catalytic performance of the material is largely influenced by the atomic structure of the material as well as any atomic defects in the structure. Therefore, finding the atomic structure and any structural defects helps develop the relation between the atomic structure and the catalytic property for further catalyst design and optimization. We will later demonstrate how the statistical approach proposed in this paper is applicable in identifying a atomic structure and structural defects. 

\begin{figure}
	\centering
	\includegraphics[width=\textwidth]{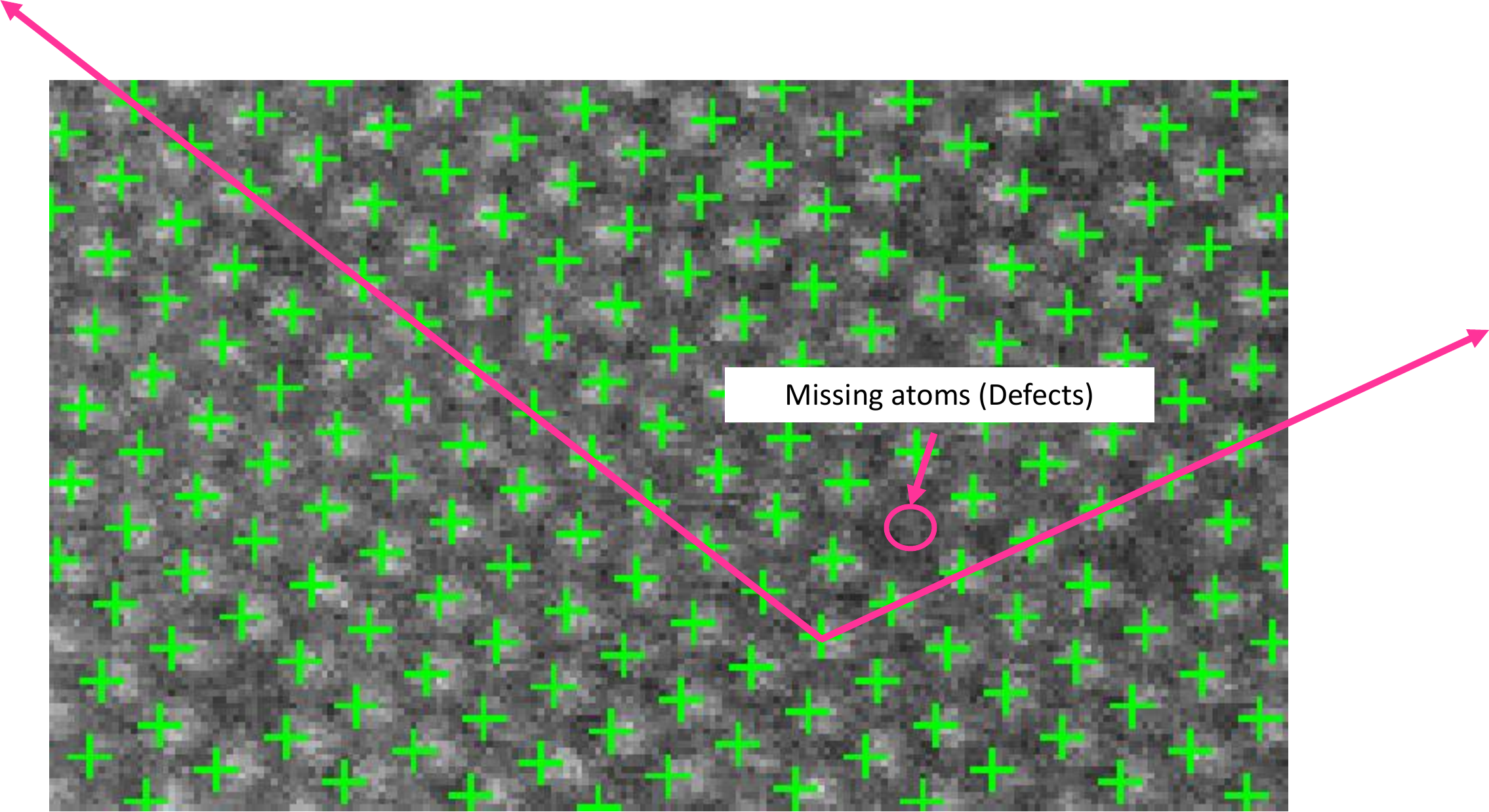}
	\caption{example atomic scale image overlaid with individual atom locations, their symmetries and defects}
	\label{fig:example}
\end{figure}

We use Figure \ref{fig:example} to illustrate a scanning transmission electron microscopy (STEM) image taken at a sub angstrom resolution and explain how an atomic structure and structural defects are defined in the image. Individual atomic columns are represented as the bright spots on a dark background. For illustrative purposes, we overlaid a green cross on the location of each atom. The reader may note that, most of the atoms lay on a regularly spaced lattice grid. The lattice grid defines the atomic structure of a sample material. However, this regularity is occasionally broken due to atomic defects. For example, a atom defect is highlighted in Figure \ref{fig:example} with a dotted magenta circle pointed out by an arrow, where an atom is supposed to exist but it is missing. The regular lattice grid and missing atoms breaking the regularity are important characteristics defining the properties of a sample material.

Current methods process atomically resolved images sequentially, first individual atoms are identified which then allow us to infer the global lattice spacing and account for defects \citep{belianinov2015identification}. The first step can be represented as a spot detection problem that locates bright spots on a dark image without a priori information. Popular approaches to spot detection include local  filtering such as the top-hat filter \citep{bright1987two} and the LoG filter \citep{sage2005automatic}. Here a filtering is applied first, and the filtered image is then thresholded to locate spots, or the h-dome method is applied to identify the local maxima of the filtered image, where an atom is then identified \citep{vincent1993morphological,smal2008new,rezatofighi2012new}. The filtering step usually relies on two or three crucial tuning parameters, including spot size, distance between spots and threshold intensity, for which some preliminary information is necessary.  However, for low contrast images having low signal-to-noise ratios, the spot detection approach is not very accurate. Besides the local filter, \cite{hughes2010likelihood} employed an approximate likelihood estimator for particle location, derived from a Poisson random field model for photon emission imaging. 

We believe that using a global lattice grid can improve the accuracy of identifying atomic locations for low contrast images, and locating atoms more accurately can improve the estimation of the global lattice. Similar to \cite{hughes2010likelihood}, the problem of estimating atom locations is formulated as a regression problem that minimizes a certain least square criterion that defines a fit to data. We add two sparsity terms to the least square criterion, a group sparsity and an individual sparsity. A choice of a specific lattice grid implies possible atomic locations restricted to the grid vertices of the chosen lattice, so selecting a lattice grid translates to selecting a representative group of atoms, which can be guided by a group sparsity penalty term. The group sparsity term penalizes selecting multiple groups since the global lattice grid is unique. On the other hand, not all atoms conform to the global lattice grid due to atomic defects. We thus utilize individual sparsity to avoid image artifacts and false positives. For the solution to approach he sparse regression formulation, we devise a group orthogonal matching pursuit algorithm with thresholding, \textit{gOMP-Thresholding}, that is scalable for large data sets and has meaningful statistical bound guarantee.  

We organize the paper as follows. Section \ref{sec:model} describes the images and the imaging physics to define a regression model, Section \ref{sec:regul} models group sparsity and individual sparsity terms and formulates the least square problem with sparsity regularization. In Sections \ref{sec:alg} and \ref{sec:detail}, we introduce the \textit{gOMP Thresholding} algorithm with proofs of performance guarantee, the implementation details and its comparison with sparse group lasso. Section \ref{sec:sim} presents numerical results of the proposed approach with simulated images. Section \ref{sec:app} demonstrates the applicability of the proposed approach for determining atomic scale material structures of important catalyst materials and identifying imaging distortions and atomic-scale lattice defects automatically. Finally, we summarize our work and and its scientific significance in Section \ref{sec:conc}.

\section{Data and Regression Model} \label{sec:model}
In this section, we describe the data and the regression model that relates structure to the imaging. Suppose that a material sample is imaged into a $M \times N$ digital image by an electron beam, and the sample consists of $T$ atoms with the $t$th atom positioned at the pixel location $(x_t, y_t)$ of the output image. Ideally, the measurement of the sample will have a sharp intensity peak at each atomic position $(x_t, y_t)$, 
\begin{equation}
f_{\delta}(x, y)= \sum_{t=1}^{T} \alpha_{t} \delta(x-x_t, y-y_t),
\end{equation}
where $\delta$ is the Dirac delta function. However, due to inherent electronics lens aberration \citep{nellist2000principles}, the STEM produces a blurred image, i.e. a convolution of the peaks with a Gaussian point spreading function $P$, 
\begin{equation}
f(x, y)= P * f_{\delta} =\sum_{t=1}^{T} \alpha_t \mbox{exp}\left(\frac{-(x-x_t)^2-(y-y_t)^2}{\tau^2} \right)
\label{conv}
\end{equation}
where $\tau^2$ is a positive constant and $*$ is the convolution operator; for the time being, we assume that $\tau^2$ is known. We do not know the number of atoms in the image and their locations $(x_t, y_t)$. Therefore, one can first pose an infinite mixture model, 
\begin{equation} 
f(x, y)=\sum_{t=1}^{\infty} \alpha_t \mbox{exp}\left(\frac{-(x-x_t)^2-(y-y_t)^2}{\tau^2} \right),
\end{equation}
and only a small number of locations can be selected by the model selection procedure. For an digital image, the infinite mixture model is equivalent to the following finite mixture model that places a mixture component at every image pixel location $(m, n)$,
\begin{equation} \label{eq:f}
f(x, y)=\sum_{(m,n) \in \mathbb{Z}_M \times \mathbb{Z}_N } \alpha_{m,n} \mbox{exp}\left(\frac{-(x-m)^2-(y-n)^2}{\tau^2} \right),
\end{equation}
where $\mathbb{Z}_M = \{1,2,\ldots, M\}$, and the value of $\alpha_{m,n}$ is 
\begin{equation*}
\alpha_{m,n}   
\begin{cases}
>0 & \text{ if  an atom exists on pixel } (m, n) \\
=0      & \text{otherwise.}
\end{cases}
\end{equation*}
We assume that the image actually measured by the STEM is a noisy version of $f$. The measured image $I(x,y)$ at pixel location $(x,y)$ is
\begin{equation} \label{sim}
I(x,y) = f(x, y) + \epsilon(x, y),
\end{equation}
where $\epsilon(x,y)$ is an independent white noise. Let $\M{Y}$ denote the $M \times N$ matrix of $I(x, y)$'s and $\M{A}$ denote the matrix of $\alpha_{m,n}$'s. We also define $\V{u}_m$ as the $M \times 1$ vector with its $i$th element equal to  $\mbox{exp}\left(-(i-m)^2/\tau^2\right)$ and $\V{v}_n$ as the $N \times 1$ vector with its $j$th element equal to $\mbox{exp}\left(-(j-n)^2/\tau^2 \right)$. The measurement then, \eqref{sim} defines the following regression model,
\begin{equation}
\M{Y} = \M{U}_{\tau} \M{A} \M{V}^T_{\tau} + \M{E},
\end{equation}
where $\M{U}_{\tau} = (\V{u}_1,....,\V{u}_M)$, $\M{V}_{\tau} = (\V{v}_1,....,\V{v}_N)$ and $\M{E}$ is the $M \times N$ noise matrix of $\epsilon(x, y)$'s. Note that unknown $\M{A}$ should be very sparse because atoms locate on a few of pixel locations. The square loss function for a choice of $\M A$ is 
\begin{equation} \label{eq:like}
L(\V A; \tau^2, \M{Y}) = ||\M{Y} - \M{U}_{\tau} \M{A} \M{V}^T_{\tau}||_F^2.
\end{equation}

\section{Structural Sparsity} \label{sec:regul}
The choice of $\V A$ can be optimized by minimizing the L2 loss $L(\V A; \tau^2, \M{Y})$ with the L1 sparsity on $\M{A}$, which is a simple spot detection problem. However, the simple spot detection does not work very well for low contrast images, resulting in many false detections and miss detections. In this section, we define a new regularization on $\M{A}$ to better guide the sparse selection of $\M{A}$ using global lattice grid information of atomic arrangements. \\

\begin{figure}
	\centering
	\includegraphics[width=0.7\textwidth]{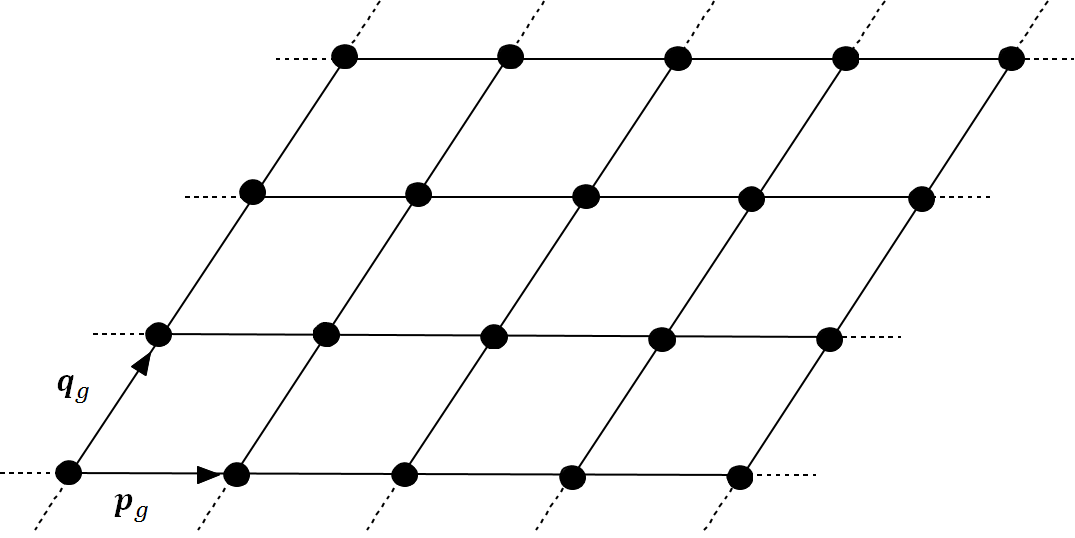}
	\caption{example Bravais lattice}
	\label{Bravais Lattice}
\end{figure}

The spatial locations of atoms in a perfect crystalline material can be completely described by a lattice group. Let $\mathbb{Z}$ denote a set of integers and $\mathbb{Z}_M = \{1,2,\ldots, M\}$. In a digital image space defined by $\mathbb{Z}_M \times \mathbb{Z}_N$, a lattice group $L_g$ is defined by  two integer-valued lattice basis $\V p_g \in \mathbb{Z}^2$ and $\V q_g \in \mathbb{Z}^2$ with the coordinate origin $\V{s}_g \in \mathbb{Z}_M \times \mathbb{Z}_N$, 
\begin{equation*}
L_{g} :=  \{ \V{s}_g + z_{p}\V{p}_g + z_{q}\V{q}_g \in \mathbb{Z}_M \times \mathbb{Z}_N; z_p, z_q \in \mathbb{Z}\},
\end{equation*}
where the subscript $g \in G$ was used to index a lattice group in a collection of possible lattice groups, and $G$ denote the collection. Figure \ref{Bravais Lattice} illustrates atoms (depicted as dots) at locations on a lattice. 
Define $\M{A}_{(g)}$ a $M \times N$ matrix with its $(m,n)$th element,
\begin{equation*}
(\M{A}_{(g)})_{m,n} = 
\begin{cases}
      \alpha_{m,n} , &\mbox{if } (m,n) \in L_{g} \\
      0, & \text{otherwise.}
    \end{cases}
\end{equation*} 
The loss function \eqref{eq:like} can be written as
\begin{equation} 
L(\V A; \tau^2, \M{Y})  = ||\M{Y} - \sum_{g \in G} \M{U}_{\tau} \M{A}_{(g)} \M{V}^T_{\tau}||_F^2.
\end{equation}
Since all atom locations belong to single lattice group (for single-crystalline material) or a few lattice groups (in the case of multi-phase materials) among all listed in $G$, we regularize $\M{A}$ with group sparsity. One possibility is to regularize $\M{A}$ with a group norm such as the group lasso penalty,
\begin{equation}
\lambda \sum_{g \in G} ||\M{A}_{(g)}||_F. 
\end{equation}
The group lasso penalty works like the lasso regularization but at a group level; where all variables in a group can shrink to zero, or all of them can be non-zero depending on $\lambda$ \citep{yuan2006model}. The group lasso criterion does not yield a within-group sparsity, i.e., the non-zero group norm $||\M{A}_{(g)}||_F$ implies all variables in group $g$ become non-zero. This is not very appropriate for our problem, because there could be some vacant locations in a chosen lattice group due to atomic defects, so some elements in the chose group could be zero. Therefore, simply regularizing $\M A$ with a group norm would result in many false atom detections. To minimize the faults, a within-group sparsity should be considered. Imposing both a group-level sparsity, and a within-group sparsity was shown previously by the sparse group lasso \citep{simon2013sparse} and the hierarchical group sparsity or more generally graph group sparsity by \cite{huang2011learning, jenatton2011structured}. In this paper, we follow \cite{huang2011learning} to develop the two-level sparsity regularization. Although the sparse group lasso provides a good alternative for our problem, it did not produce good results with the sparse group lasso due to several reasons discussed in more details in Section \ref{sec:SGL}. 

Consider a set of the lattice groups, $\{L_g; g \in G\}$, and define $S_{m,n}$ as a group of singleton $(m,n)$. The singleton groups and the lattice groups form the following inclusion relation: for each $S_{m,n}$, there exists $g$ that satisfies
\begin{equation*}
S_{m,n} \subset L_g.
\end{equation*}
In addition, $L_g$ and the entire image $\mathbb{Z}_M \times \mathbb{Z}_N$ has the relation $L_g \subset \mathbb{Z}_M \times \mathbb{Z}_N$. These inclusion relations can be represented by a tree hierarchy that has $\mathbb{Z}_M \times \mathbb{Z}_N$ as a root node, all $L_g$'s as the first level children, and all $S_{m,n} \subset L_g$ as the second level children of $L_g$. Following \citet[Section 4.3]{huang2011learning}, we represent the sparsity cost for the tree as 
\begin{equation}
C(\V A) = \left\{\log_2 (2|G|) \cdot \sum_{g \in G} ||tr(\M{A}_{(g)}\M{A}_{(g)}^T)||_0 + \log_2 2K \cdot ||\M{A}||_0\right\},
\label{c_def}
\end{equation}
where $K = \max_{g \in G} |L_g|$. Note that the first cost term represents the group-level sparsity, and the second term represents the within-group sparsity.  We find $\M A$ that solves 
\begin{equation} \label{eq:plike}
\mbox{Minimize } \quad L(\V A; \tau^2, \M{Y}) \mbox{ subject to } C(\V A) \le c,
\end{equation}
where $c > 0$ is a tuning parameter. The cost of making group $\M{A}_{(g)}$ nonzero is $\log_2 (2|G|)$, which is much smaller than the cost for adding an individual, $\log_2 2K$, and the criterion favors a group selection unless there are strong counter evidences from $L(\V A; \tau^2, \M{Y})$. The problem is a non-convex optimization problem, and a sub-optimal solution will be pursued. 

\section{Group OMP with Thresholding} \label{sec:alg}
Problem \eqref{eq:plike} involves minimizing the square loss $L(\V A; \tau^2, \M{Y})$ with structural sparsity regularization $C(\M A)$. \citet{huang2011learning} proposed an heuristic approach to solve a general structural sparsity regularization problem, which is applicable when the structural sparsity term originates from tree structured or graph structured groupings of data elements. The algorithm can be directly applicable to our problem since $C(\M A)$ originated from a tree structured grouping of atom locations. However, doing so is computationally inefficient. We revise the algorithm for efficiency and also show that the consistency and convergence results for the original algorithm are still applicable for the modification.

The group orthogonal matching pursuit (gOMP) algorithm proposed by \citet{huang2011learning} solves the regularized regression problem that minimizes a square loss under a structural sparsity regularization. The algorithm iteratively selects a group or an individual variable in each iteration that improves its square loss most while bounding the regularization term below a certain threshold. The number of iterations can be equal to the number of the non-zero elements in groundtruth $\V A$ for the worst case. For example, in the groundtruth all non-zero variables belong to a group but some variables in the group are zero. In this case, the group selection does not fully explain the groundtruth and many individual selections have to be performed, yielding many iterations. We revise the algorithm by splitting the variable selection iteration into two levels, group-level selection and within-group selection. Each iteration first selects a group of variables and then applies a marginal regression to choose non-zero elements within the chosen group. Algorithm \ref{alg:gOMP} describes the details of the algorithm. Let $g_k$ denote the index of the lattice group selected at iteration $k$ and also denote $F^{(k)} = \cup_{l = 1}^k L_{g_l}$ and 
\begin{equation*}
\V{\hat A}^{(k)} = \arg\min L(\V A; \tau^2, \M{Y}) \mbox{ subject to } \mbox{supp(}\V A) \subset F^{(k)},
\end{equation*}
where $\mbox{supp(}\V A) = \{(m, n); (\M A)_{m,n} \neq 0 \}$. For the group selection, we follow  \citet{huang2011learning} to select $g_k \in G$ that maximizes the following gain ratio,
\begin{equation} \label{eq:gratio}
\phi(g_k) = \frac{L(\M{\hat{A}}^{(k-1)}; \tau^2, \M{Y}) - L(\M{\hat{A}}^{(k)}; \tau^2, \M{Y}) }{ C(\M{\hat{A}}^{(k)}) - C(\M{\hat{A}}^{(k-1)}) }.
\end{equation}

The group selection augments a set of non-zero elements (of $\M A$) from $F^{(k-1)}$ to $F^{(k)} = L_{g_k} \cup F^{(k-1)}$. Then, we apply the marginal regression \citep{genovese2012comparison} for a sparse solution of the following regression model,
\begin{equation*}
\M{Y} = \V{U}_{\tau} \M A \V{V}_{\tau}^T + \M{E} \mbox{ subject to } \mbox{supp(}\V A) \subset F^{(k)}.
\end{equation*}
For the marginal regression, first compute the marginal regression coefficients,
\begin{equation*}
\left(\M{\hat{A}}^{(k)}_{\gamma} \right)_{m,n} = \V{u}^T_{m}  \M{Y} \V{v}_{n} \mbox{ if } (m, n) \in F^{(k)} \mbox{ or 0 otherwise.}
\end{equation*}
Each of the marginal regression coefficients is thresholded by a threshold  $\rho$, 
\begin{equation*}
\left(\V{\hat A}^{(k)}_{\rho} \right)_{m,n} = \left(\V{\hat A}^{(k)}_{\gamma} \right)_{m,n} 1\left\{\left(\V{\hat A}^{(k)}_{\gamma} \right)_{m,n} \ge \rho \right\}.
\end{equation*}
The iterations of the group selection step and the subsequent marginal regression are repeated as long as a sparsity condition $C(\V{\hat A}^{(k)}_{\rho}) \le c$ is satisfied. We call the whole algorithm  $gOMP$-$Thresholding$. The proposed algorithm has multiple tuning parameters, the list of potential lattice groups $\{L_g; g \in G\}$, bandwidth $\tau^2$ for a point spreading function, constant $c$ that defines the stopping condition of the iterations, and threshold $\rho$ that defines the within-group sparsity. The choices of the tuning parameters and the convergence and statistical analysis with the choices will be presented in the next section. 
\begin{algorithm}
\caption{$gOMP$-$Thresholding$} \label{alg:gOMP}
\begin{algorithmic}[1]
\REQUIRE parameter $\tau^2$, the list of potential groups $\{ L_g; g \in G\}$, stopping criterion $c$, threshold $\rho$
\INPUT   input image $\M{Y}$
\OUTPUT  $\M{A}$
\STATE \textbf{Initialization} $F^{(0)}=\emptyset$ and $\hat{\V A}^{(k)}_{\rho}=\V 0$
\WHILE{$C(\M{\hat A}^{(k)}_{\rho}) < c$}
	\STATE $k = k + 1$
	\STATE Select $g_k \in G$ to maximize $\phi(g_k)$ following the group selection criterion \eqref{eq:gratio}.
	\STATE Let $F^{(k)}=L_{g(k)} \cup F^{(k-1)}$.
	\STATE Marginal Regression: $\left(\M{\hat A}^{(k)}_{\gamma} \right)_{m,n} = \V{u}^T_{m}  \M{Y} \V{v}_{n}$ if $(m, n) \in F^{(k)}$ or zero otherwise.
	\STATE Thresholding: $\left(\M{\hat A}^{(k)}_{\rho} \right)_{m,n} = \left(\M{\hat A}^{(k)}_{\gamma} \right)_{m,n} 1\left\{\left(\M{\hat A}^{(k)}_{\gamma} \right)_{m,n} \ge \rho \right\}$.
\ENDWHILE
\end{algorithmic}
\end{algorithm}

\section{Implementation Details} \label{sec:detail}
This section contains information on choosing the tuning parameters of the proposed algorithm. 
\subsection{Listing $L_g$'s and Estimating $\tau$} \label{sec:G}
The proposed approach requires the list of the lattice groups $L_g$ that may be found in the input image. Certainly, one can consider all possible lattice groups with all possible combinations of $\V s_g$, $\V p_g$ and $\V q_g$. However, the number is theoretically infinite or could be very large even when only finite numbers of uniformly sampled values of $\V s_g$, $\V p_g$ and $\V q_g$ are considered. Fixing $\V p_g$ and $\V q_g$ to estimates allows us to narrow down the number of the possible lattice groups to the range of $\V s_g$. Let $\V{\hat{p}}$ and $\V{\hat{q}}$ denote these estimates. Due to the lattice periodicity, the range of possible $\V{s}_g$ can be restricted to the parallelogram formed by two basis vectors $\V{\hat{p}}$ and $\V{\hat{q}}$,
\begin{equation*}
\V{s}_g \in \{ a_p \V{\hat{p}} + a_q \V{\hat{q}} \in \mathbb{Z}_M \times \mathbb{Z}_N; a_p, a_q \in [0, 1) \}.
\end{equation*}
Where $\times$ is a cross product operator of two vectors. Note that $||\V{\hat{p}} \times \V{\hat{q}}||_2$ is the area of the parallelogram formed by two basis vectors $\V{\hat{p}}$ and $\V{\hat{q}}$, which is equal to the number of pixel locations in the parallelogram. Since the four vertices of the parallelogram represent the same $\V s_g$ due to the lattice periodicity, the number of all possible $\V{s}_g$ is the area minus redundancy, i.e., $||\V{\hat{p}} \times \V{\hat{q}}||_2-3$. Using $\V{\hat{p}}$ and $\V{\hat{q}}$ with the range of $\V{s}_g$, we can list all possible lattice groups, and the $g$th group as
\begin{equation*}
L_g = \{ \V{s}_g + z_{p}\V{\hat{p}} + z_{q}\V{\hat{q}} \in \mathbb{Z}_M \times \mathbb{Z}_N; z_p, z_q \in \mathbb{Z}\}.
\end{equation*}
Note that $L_g \bigcap L_{g'} = \emptyset$ and $\bigcup_{g \in G} L_g = \mathbb{Z}_M \times \mathbb{Z}_N$. The groups form a non-overlapping  partition of $\mathbb{Z}_M \times \mathbb{Z}_N$. In the remainder of this section, we discuss strategies for good estimates of $\V{\hat{p}}$ and $\V{\hat{q}}$ under low signal-to-noise ratio and the case of missing atoms.

\begin{figure}
	\centering
		\includegraphics[width=\textwidth]{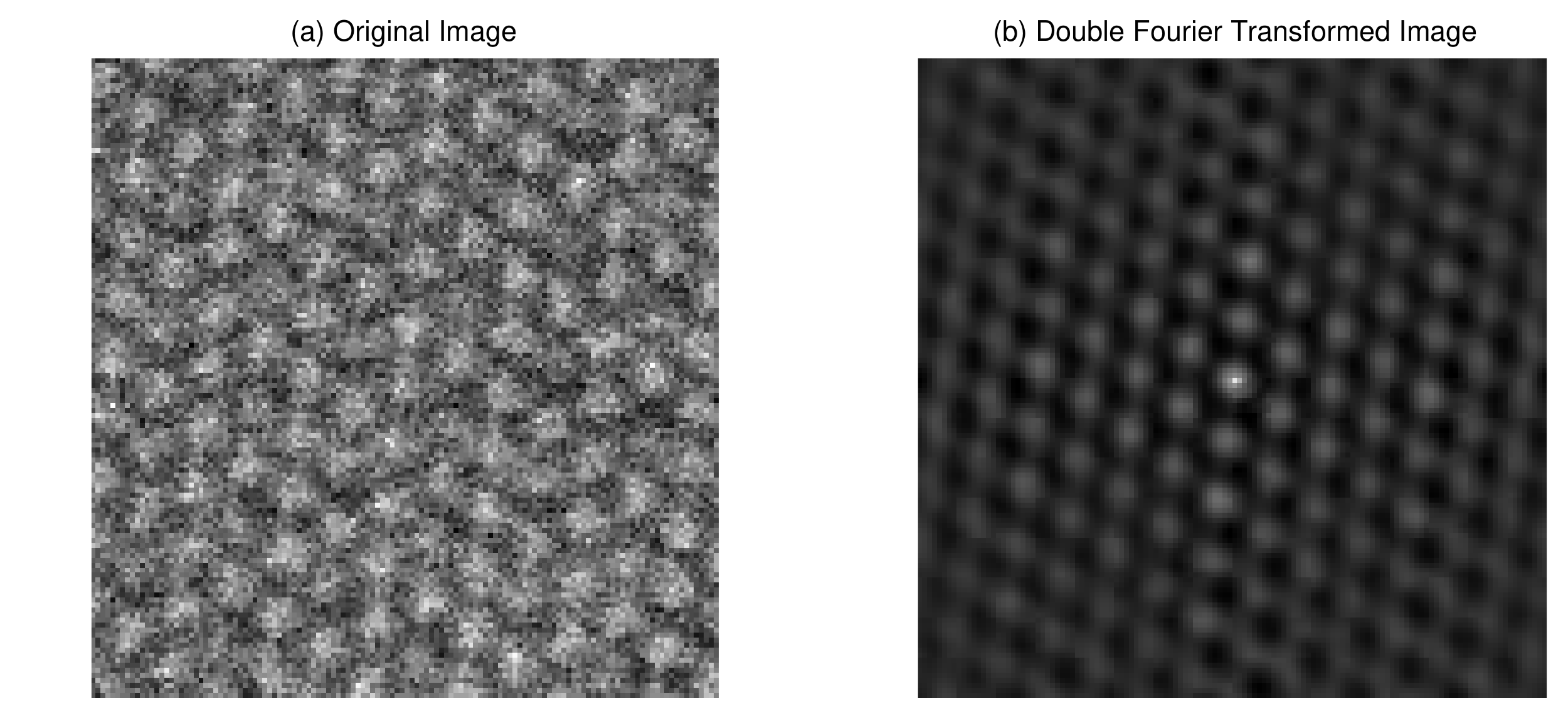}
	\caption{comparison of (a) an example input image and (b) its double fourier transform}
	\label{stem}
\end{figure}

Estimating two basis vectors $\V{\hat{p}}$ and $\V{\hat{q}}$ is difficult partially due to low contrast of an input image and partially because the input image does not contain a perfect lattice. We will use the double fourier transform of an input image $I$ to achieve the estimates $\V{\hat{p}}$ and $\V{\hat{q}}$. The double fourier transform is defined by the fourier transform of the square of the fourier transform of $I$,
\begin{equation*}
\mathcal{F} \{ |\mathcal{F}\{I\}|^2 \},
\end{equation*}
where $\mathcal{F}$ is a 2d fourier transform operator. According to \eqref{sim}, an input image is 
\begin{equation*}
I(\V{x}) = f(\V{x}) + \epsilon(\V{x}),
\end{equation*}
where $\V{x} = (x, y)$ denotes an two-dimensional image coordinate. The main signal $f(\V{x})$ is the contribution by all atoms on the underlying lattice minus the contribution by missing atoms, 
\begin{equation*}
f(\V{x}) = P * f_{\delta}(\V{x}) \\
         = P * \left( \sum_{\V{x}_l \in L_g} \alpha_{\delta} \delta(\V{x} - \V{x}_l) - \sum_{\V{x}_e \in E_g} \alpha  \delta(\V{x} - \V{x}_e) \right),
\end{equation*}
where $E_g \subset L_g$ is the set of the locations where atoms are missed. Let $f^*_{\delta} = \sum_{\V{x}_l \in L_g} \alpha_{\delta} \delta(\V{x} - \V{x}_l)$ and $e^*_{\delta} = \sum_{\V{x}_e \in E_g} \alpha  \delta(\V{x} - \V{x}_e)$. The  fourier transform of the input image is
\begin{equation*}
\begin{split}
\mathcal{F}\{I\} = \mathcal{F}\{P\} \mathcal{F}\{f^*_{\delta}\} - \mathcal{F}\{P\} \mathcal{F}\{e^*_{\delta}\}  + \mathcal{F} \{\epsilon\}.
\end{split}
\end{equation*}
Typically, the cardinality of $E_g$ is ignorably small compared to the cardinality of $L_g$, and the locations in $E_g$ are randomly distributed over the entire space of an input image. Since the power spectrum of a signal with randomly locating peaks is uniform, and the uniform magnitude is proportional to the cardinality of $E_g$, the effects of $\mathcal{F}\{P\} \mathcal{F}\{e^*_{\delta}\}$ on the total fourier coefficient is ignorable,
\begin{equation*}
\begin{split}
\mathcal{F}\{I\} \approx \mathcal{F}\{P\} \mathcal{F}\{f^*_{\delta}\}  + \mathcal{F} \{\epsilon\}.
\end{split}
\end{equation*}
Assume that the fourier transform of noise and the fourier transform of signal are nearly orthogonal, which is true for many practical cases since the noise is typically described by high frequency components and the signal is mostly described by low frequency components. Therefore, $\mathcal{F}\{f^*_{\delta}\} \mathcal{F} \{\epsilon\} \approx 0$, and we have
\begin{equation*}
\begin{split}
|\mathcal{F}\{I\}|^2 = |\mathcal{F}\{P\}|^2|\mathcal{F}\{f^*_{\delta}\}|^2 + |\mathcal{F} \{\epsilon\}|^2.
\end{split}
\end{equation*}
The double fourier transform of the input image is the fourier transform of $|\mathcal{F}\{I\}|^2$, 
\begin{equation} \label{eq:dft}
\begin{split}
\mathcal{F} \{|\mathcal{F}\{I\}|^2 \} = \mathcal{F}\{|\mathcal{F}\{P\}|^2 \} * \mathcal{F}\{|\mathcal{F}\{f^*_{\delta}\}|^2\} + \mathcal{F} \{|\mathcal{F} \{\epsilon\}|^2\}.
\end{split}
\end{equation}
Since the fourier transform of a Gaussian point spread function $P$ is a Gaussian point spread function, $|\mathcal{F}\{P\}|^2$ is a Gaussian point spread function with $\sqrt{2}$ times wider spreading width than original, and so is $\mathcal{F}\{|\mathcal{F}\{P\}|^2 \}$, which we denote by $\tilde{P}$. Since the fourier transform is orthonormal transformation, the real parts and the imaginary parts of the fourier transform of a Gaussian white noise are Gaussian white noises, so $|\mathcal{F} \{\epsilon\}|^2$ is a constant multiple of a (non-centered) chi-square random variable with degree 2, and $\mathcal{F}\{|\mathcal{F} \{\epsilon\}|^2\}$ is a linear combination of (non-centered) chi-square random variables, which we denote by $\tilde{\epsilon}$; $\tilde{\epsilon}$ is still independent white noises since the fourier transform is orthonormal. Therefore, we can simplify the previous expression to
\begin{equation*}
\begin{split}
\mathcal{F} \{|\mathcal{F}\{I\}|^2 \} = \tilde{P} * \mathcal{F}\{|\mathcal{F}\{f^*_{\delta}\}|^2\}  + \tilde{\epsilon}.
\end{split}
\end{equation*}
The fourier transform of  $f^*_{\delta}$ is 
\begin{equation*}
\begin{split}
\mathcal{F} \{f^*_{\delta}\}(\V{u}) &= \sum_{\V{x}_l \in L_g} \alpha_{\delta} \int \delta(\V{x} - \V{x}_l) \exp\{- j \V{u}^T \V{x}\} \\
                                      &= \sum_{\V{x}_l \in L_g} \alpha_{\delta} \exp\{- j \V{u}^T \V{x}_l\},
\end{split}
\end{equation*}
and its square is
\begin{equation*}
\begin{split}
|\mathcal{F} \{f^*_{\delta}\}(\V{u})|^2 &= \alpha_{\delta}^2 \left(\sum_{\V{x}_l \in L_g} \cos( \V{u}^T \V{x}_{l})\right)^2 + \alpha_{\delta}^2 \left(\sum_{\V{x}_l \in L_g} \sin( \V{u}^T \V{x}_{l})\right)^2.
\end{split}
\end{equation*}
Note $\V{x}_l \in L_g$ is represented by 
\begin{equation}
\V{x}_l = \V{s}_{g} + z_{p, l} \V{p}_g + z_{q, l} \V{q}_g \mbox{ for } z_{p,l}, z_{q, l} \in \mathbb{Z}.
\end{equation}
The square of the fourier transform is simplified to
\begin{equation*}
\begin{split}
|\mathcal{F} \{f^*_{\delta}\}(\V{u})|^2
&= \alpha_{\delta}^2  \left(\sum_{z_{p,l}, z_{q, l} \in \mathbb{Z}} \cos( \V{u}^T (\V{s}_{g} + z_{p, l} \V{p}_g + z_{q, l} \V{p}_g))\right)^2 \\
& +\alpha_{\delta}^2  \left(\sum_{z_{p,k}, z_{q, k} \in \mathbb{Z}} \sin( \V{u}^T (\V{s}_{g} + z_{p, k} \V{p}_g + z_{q, k} \V{p}_g))\right)^2 \\
&= \alpha_{\delta}^2 \sum_{z_{p,l}, z_{q, l},z_{p,k}, z_{q, k}} 1+ 2\cos( \V{u}^T ((z_{p,l} - z_{p,k}) \V{p}_g + (z_{q,l} - z_{q,k}) \V{q}_g))\\
& = \alpha_{\delta}^2 \sum_{z_{p}', z_{q}' \in \mathbb{Z}}\left(\frac{1}{z_{p}'z_{q}'}\right)(1+ 2\cos( \V{u}^T (z_{p}' \V{p}_g + z_{q}' \V{q}_g))).
\end{split}
\end{equation*}
Let $\V{\tilde{x}}_l = z_{p}' \V{p}_g + z_{q}' \V{q}_g$ and $\tilde{L}_g = \{z_{p}' \V{p}_g + z_{q}' \V{q}_g;z_{p}', z_{q}' \in \mathbb{Z}\}$. The previous expression for $|\mathcal{F} \{f^*_{\delta}\}(\V{u})|^2$ can be written as
\begin{equation*}
|\mathcal{F} \{f^*_{\delta}\}(\V{u})|^2 \propto \sum_{\V{\tilde{x}}_l \in \tilde{L}_g}\frac{1}{||\V{\tilde{x}}_l||_2} (1+ 2\cos( \V{u}^T \V{\tilde{x}}_l)).
\end{equation*}
Using the result, we can derive the double fourier transform of $f^*_{\delta}$,
\begin{equation*}
\begin{split}
\mathcal{F}\{|\mathcal{F} \{f^*_{\delta}\}|^2 \}(\V \omega) \propto \sum_{\V{\tilde{x}}_l \in \tilde{L}_g} \frac{1}{||\V{\omega}||_2} \delta(\V{\omega}- \V{\tilde{x}}_l).
\end{split}
\end{equation*}
Therefore, the double fourier transform image of $\M{Y}$ is approximately
\begin{equation*}
\begin{split}
\mathcal{F} \{|\mathcal{F}\{I\}|^2 \}(\V \omega) = h \tilde{P} * \sum_{\V{\tilde{x}}_l \in \tilde{L}_g} \frac{1}{||\V{\omega}||_2} \delta(\V{\omega}- \V{\tilde{x}}_l)  + \tilde{\epsilon}.
\end{split}
\end{equation*}
where $h$ is an constant. Note that the double fourier transform has peaks spaced every $\V{\tilde{x}}_l \in \tilde{L}_g$, and the $\tilde{L}_g$ has the exactly same basis vectors as the original lattice group $L_g$ of the input image and it is invariant to any spatial shift $\V{s}_g$ of the lattice locations. In addition, the peaks in the double fourier transform are much more amplified in lower frequency bands that corresponds to smaller $||\V{\omega}||_2$ while the noise $\tilde{\epsilon}$ is still independently and identically distributed over $\V{\omega}$. Therefore, the lower frequency region of the double fourier transform reveals the original lattice basis vectors with a very high SNR ratio.  Figure \ref{stem} show an example image and its DMFT, which is consistent with the pattern.
We could apply an existing spot detection algorithm, in particular, the determinant of the Hessian \citep{bay2008speeded}, on the double fourier transform image to estimate $\V p_g$ and $\V q_g $ and also estimate the spreading width of the point spreading function $\tilde{P}$ on the double fourier transform domain, which we denote by $\tilde{\tau}$. According to our derivation \eqref{eq:dft} and the subsequent discussion, the spreading width $\tilde \tau$ is $\sqrt{2}$ times wider than the spreading width $\tau$ of the point spreading function $P$ in the input image. Therefore, once $\tilde \tau$ is estimated, $\tau \approx \frac{1}{\sqrt{2}} \tilde \tau$. 

\subsection{Choice of Stopping Condition Constant $c$ and Related Error Bounds} \label{sec:stop}
Let $\M{\bar A}$ denote the true signal to estimate. We first present the error bound of the solution of the proposed algorithm to the groundtruth, for which we introduce some notations. For all $F\subset \mathbb{Z}_M \times \mathbb{Z}_N$, define
\begin{equation*}
\eta_{+}(F)=\mbox{sup}\left\{\frac{1}{MN}||\M U_{\tau}\M A\M V_{\tau}^T||_F^2/||\V{A}||_F^2:\ \mbox{supp}(\V{A})\subset F\right\}, \mbox{ and}
\end{equation*}
\begin{equation*}
\eta_{-}(F)=\mbox{inf}\left\{\frac{1}{MN}||\M U_{\tau}\M A\M V_{\tau}^T||_F^2/||\V{A}||_F^2:\ \mbox{supp}(\V{A})\subset F \right\}.
\end{equation*}
Moreover, for $c>0$, define
\begin{equation*}
\begin{split}
& \eta_{+}(c)=\mbox{sup}\{\eta_{+}(\mbox{supp}(\M A)): C(\M A)<c\}, \\
& \eta_{-}(c)=\mbox{inf}\{\eta_{-}(\mbox{supp}(\M A)): C(\M A)<c\}, \mbox{ and} \\
& \eta_0 = \mbox{sup} \{\eta_{+}(L_g): g\in G\}. % \\
%& C_0    = \mbox{sup} \{C(\M A_{(g)}); g \in G \}.
\end{split}
\end{equation*}

\begin{theorem}
Consider the true signal $\V{\bar A}$ and $\epsilon$ such that 
$$\epsilon\in(0,||\M Y||_F^2-||\M{Y} - \M{U}_{\tau} \M{\bar{A}} \M{V}_{\tau}^T||_F^2].$$ 
If the choice of $c$ satisfies
$$c\geq\frac{\eta_0 C(\M{\bar A)}}{\nu\eta_{-}(c+C(\M{\bar A}))}\log\frac{||\M Y||_F^2-||\M{Y} - \M{U}_{\tau} \M{\bar{A}}  \M{V}_{\tau}^T||_F^2}{\epsilon} \mbox{ for } \nu \in (0, 1],$$ 
with probability $1-p$, 
\begin{equation*}
||\M{\hat A}^{(k)}_{\rho} - \V{\bar A}||_F^2 \le \frac{10 ||\M U_{\tau} \V{\bar A} \M V_{\tau}^T - \mathbb{E}[\M{Y}]||_F + 37\sigma^2 (c + \eta_0)+ 29\sigma^2 \log(6/p) + 2.5\epsilon}{MN\eta_{-} (c + C_0 + C(\M{\bar A}))}
\end{equation*}
%$$||\M{Y} - \M{U}_{\tau} \M{\hat A}^{(k)}_{\rho} \M{V}_{\tau}^T||_F^2\leq ||\M{Y} - \M{U}_{\tau} \M{\bar{A}} \M{V}_{\tau}^T||_F^2+\epsilon.$$
\label{c_tuning}
\end{theorem} 
The proof of Theorem \ref{c_tuning} is straightforward using \citet[Theorems 6 and 9]{huang2011learning}. The theorem implies that the $gOMP$-$Thresholding$ output $\M{\hat A}^{(k)}_{\rho}$ is within the stated error bound to the groundtruth $\bar{\V A}$ with a proper choice of $c$ that satisfies the condition stated in the theorem. However, the theorem does not provide any practical guidance in how to choose $c$ because the condition for $c$ is not computable. 

On the other hand, the choice of $c$ is related to the number of the lattice groups selected to describe an input image, since every iteration of the proposed algorithm selects exactly one lattice group and $c$ determines the number of iterations to run. When the number of lattice groups existing in the input image is known, the number can be used to determine $c$. For example, when single crystalline material is imaged, there is only one lattice type with basis vectors $\V p_g$ and $\V q_g$. The number of atoms on the lattice group within a $M \times N$ digital image is $\frac{MN}{||\V p_g \times \V p_g||}$. The stopping condition $c$ should be set to the cost $C(\cdot)$ for one group selection, 
\begin{equation*}
c = \log_2 (2|G|) + \log_2 (2K) \frac{MN}{||\V p_g \times \V p_g||},
\end{equation*}
which is applied for all of our numerical experiments.

\subsection{Choice of Threshold $\rho$} \label{sec:threshold}
The threshold parameter $\rho$ is applied on the marginal regression outcome $\M{\hat A}^{(k)}_{\gamma}$ for a within-group sparsity. In this section, we describe how to choose this parameter. Let $\rho_j$ denote the value of the $j$th largest element of $\hat{\V A}^{(k)}_{\gamma}$. Define
\begin{equation}
S_j = \{(m, n) \in \mathbb{Z}_M \times \mathbb{Z}_N: \left(\hat{\V A}^{(k)}_{\gamma} \right)_{m,n} \ge \rho_j \}.
\end{equation}
In addition, let $\V{y} = vec(\M Y)$, let $\M X_j$ denote $MN \times |S_j|$ matrix with columns $\{\V v_n \otimes \V u_m: (m,n) \in S_j \}$, where $\otimes$ denotes a Kroneck product, and let $\M{\Sigma}_j = \M X_j^T \M X_j$ and $\M H_j = \M X_j (\M X_j^T \M X_j)^{-1}\M X_j^T$. Similarly, when $\bar S = \mbox{supp}(\M{\bar A})$ denote the support of the ground truth solution, $\M{\bar X}$ denote $MN \times |\bar S|$ matrix with columns $\{\V v_n \otimes \V u_m: (m,n) \in \bar S \}$ and $\M{\bar \Sigma} =  \M{\bar X}^T \M{\bar X}$. 
\begin{theorem}
Let $\bar{a} = \inf\{(\M{\bar A})_{m,n}: (m,n) \in \bar S\}$ and $\bar b =\inf\{||\M{\bar\Sigma} \V \mu||_2: ||\V \mu||_2=1 \}$. For $q \geq 1$, set 
\begin{equation}
 j^* = \min\{j: Del(j) < \sigma\delta_k \}.
 \label{rho_tuning}
 \end{equation}
 where $Del(j) = ||(\M H_{j+1} -\M{H}_j)\V y||_F^2$, $\delta_k = q\sqrt{2 \log ||\M{\hat A}^{(k)}_{\gamma}||_0}$, and $\sigma$ is the standard deviation of a Gaussian observation noise in $\M{Y}$. If the following condition holds
\begin{equation}
\bar a \geq 2q\sigma\bar b^{-1/2}\sqrt{\frac{2\log ||\M{\hat A}^{(k)}_{\gamma}||_0}{MN}} \mbox{ and } \bar S \subset \mbox{supp}(\M{\hat A}^{(k)}_{\gamma}),
\end{equation}
then $S_{j^*} = \bar S$ with probability no le ss than $1-4 \log ||\M{\hat A}^{(k)}_{\gamma}||_0^{-q^2}$. 
\label{rho_theorem}	
\end{theorem}
The proof of the theorem  is straightforward using \citet[Theorem 5]{slawski2013non}. Theorem \ref{rho_theorem} provides the statistical guarantee of support recovery for the proposed $gOMP$-$Thresholding$ algorithm with the choice of threshold $\rho_{j^*}$. Getting $j^*$ requires the noise level $\sigma$ which is unknown. In practice, we can use the naive plug-in estimation of $j^*$,
\begin{equation} \label{eq:thresh}
\hat{j}^* =  \min\{j: Del(j) < \hat\sigma\delta_k\},
\end{equation}
where $\hat{\sigma}^2 = \frac{||\M{Y} - \V{U}_{\tau}\M{\hat A}^{(k)}_{\gamma}\V{V}_{\tau}^T||_F^2}{MN-1}$. The corresponding threshold estimate is $\rho_{\hat{j}^*}$. The naive estimation yielded satisfactory results for all our numerical examples. The practical meaning of $j$ is the number of atoms in image $\M Y$, and $\hat{j}^*$ implies the estimate of that number. 

\subsection{Comparison to the Sparse Group Lasso} \label{sec:SGL}
As an alternative to the proposed approach, one can consider the following sparse group lasso (SGL) formulation,
\begin{equation*}
\mbox{Minimize} \quad ||\M{Y} - \sum_{g \in G} \M{U}_{\tau} \M{A}_{(g)} \M{V}_{\tau}^T||_F^2 + \lambda_1 \sum_{g \in G} ||\M{A}_{(g)}||_F + \lambda_2 ||\V A||_1
\end{equation*}
where $||\V A||_1=\sum_{m=1}^M\sum_{n=1}^{N}|(\V A)_{m,n}|$ is the elementwise L1 norm of $\V A$. \cite{chatterjee2012sparse} showed that the SGL regularizer is a special case of regularization with the hierarchical tree induced sparsity norm \citep{liu2010moreau,jenatton2011structured} that we applied for our formulation, and provided explicit bounds for the consistency of the SGL. \cite{liu2010moreau} proposed a sub-gradient approach to solve the SGL problem. In this section, we compare the SGL with our proposed approach numerically. 

\begin{figure}
	\centering
	\includegraphics[width=0.5\textwidth]{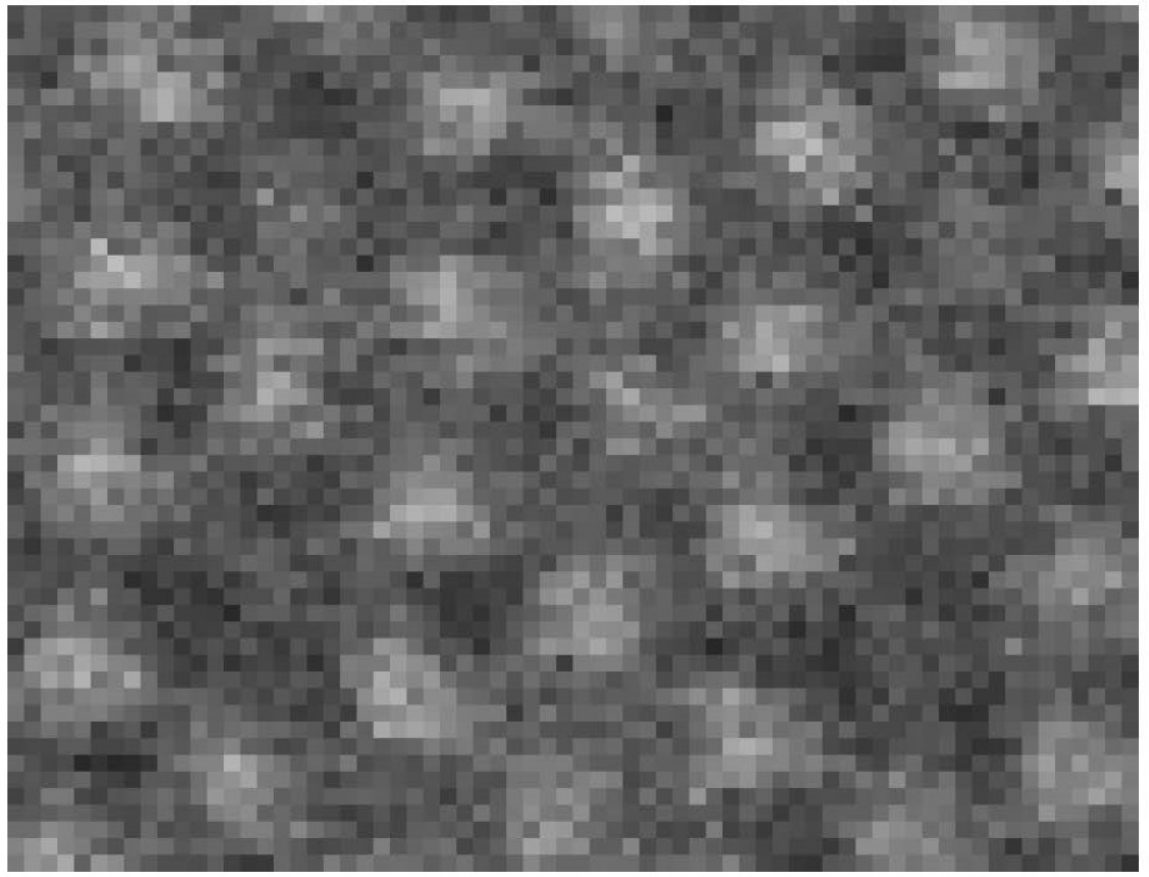}
	\caption{test image}
	\label{Truth_Compare}
\end{figure}

\begin{figure}
	\centering
	\begin{subfigure}{0.45\textwidth}
		\includegraphics[width=\textwidth]{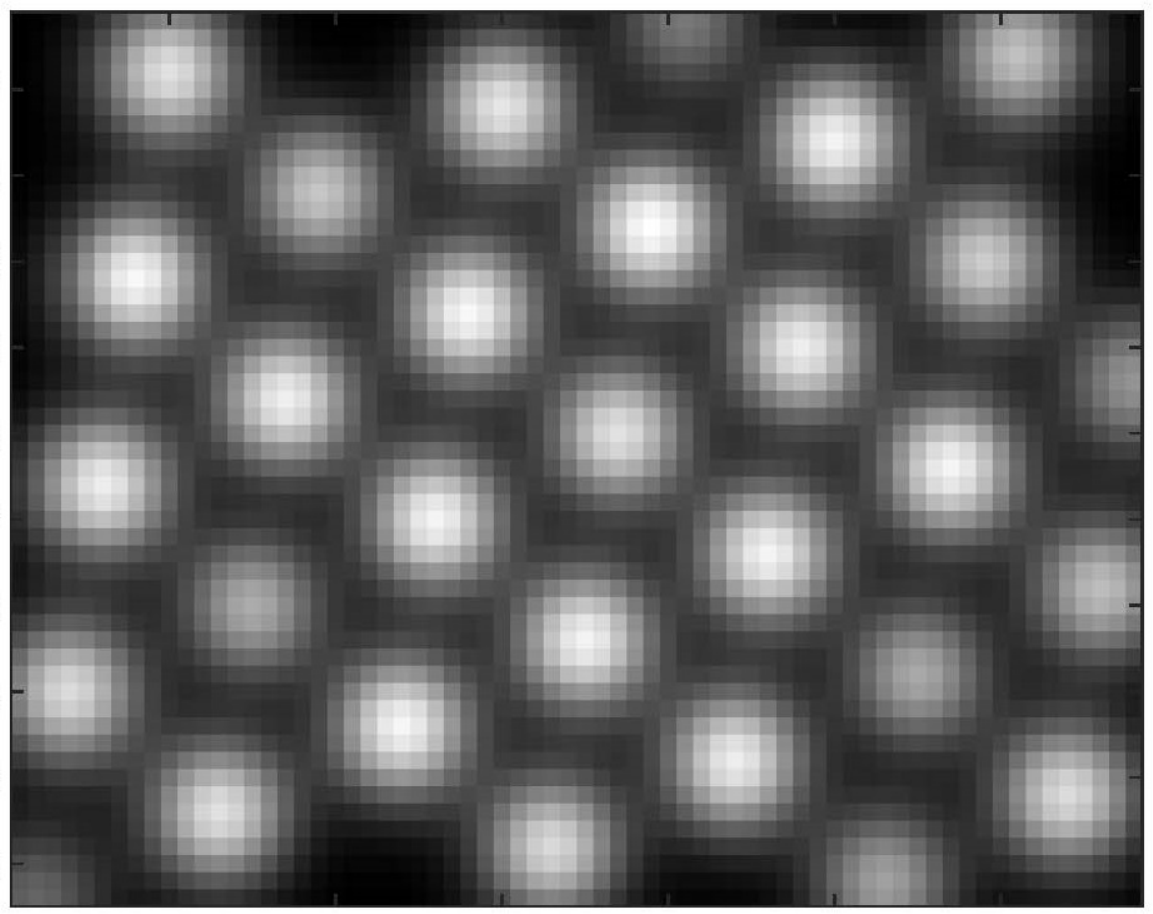}
		\caption{}
	\end{subfigure}
	\begin{subfigure}{0.45\textwidth}
		\includegraphics[width=\textwidth]{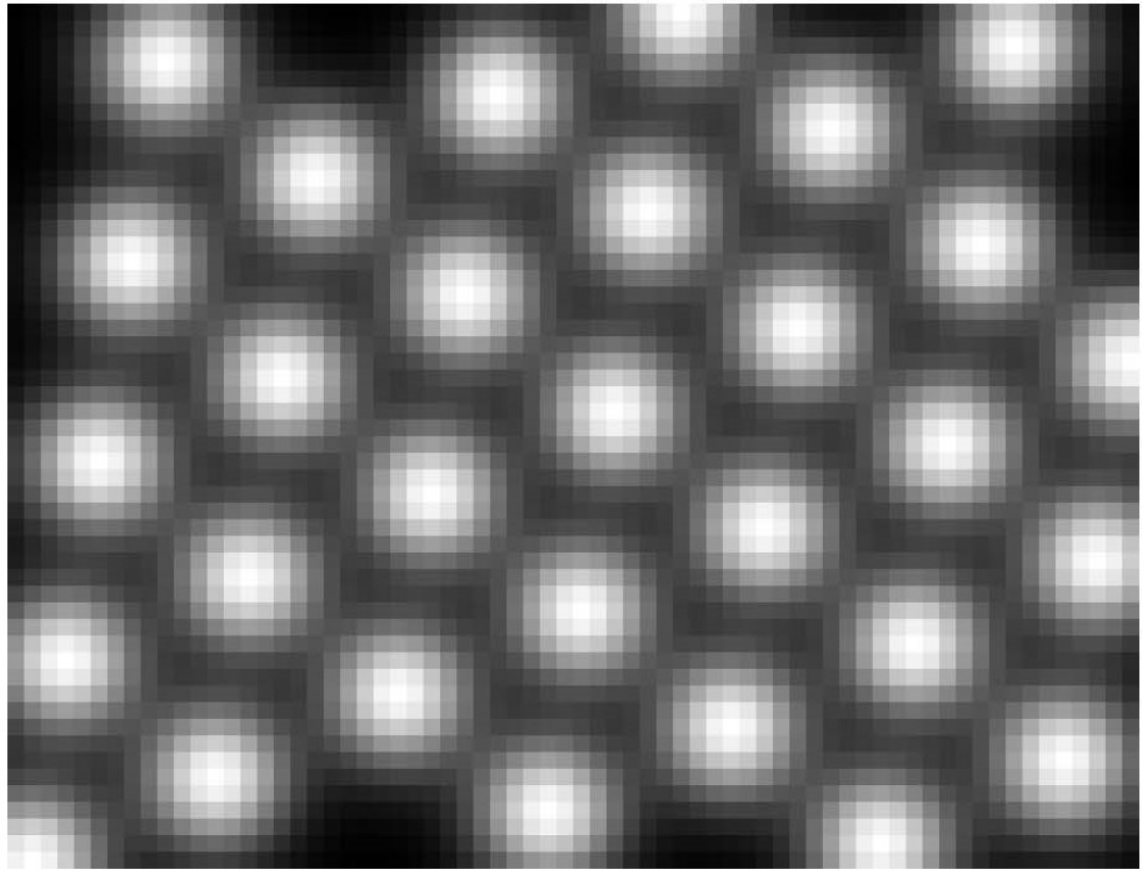}
		\caption{}
	\end{subfigure}
	\caption{comparison of the sparse group lasso (SGL) and the $gOMP$ with no thresholding. (a) SGL with $\lambda_2$ fine-tuned while fixing $\lambda_1$ to a small constant. (b) $gOMP$ with no thresholding step.}
	\label{Reconstruct_Compare}
\end{figure}

We used a test image shown in Figure \ref{Truth_Compare}, where all atoms belong to one lattice, but there are missing locations. We used the MATLAB package named \texttt{SLEP} for the SGL that implements the sub-gradient algorithm proposed by \citet{liu2010moreau}, where tuning penalty parameters $\lambda_1$ and $\lambda_2$ proved crucial to achieve good results. Performing a popular cross-validation selection that exhaustively searches the two dimensional space of $(\lambda_1,\lambda_2)$ is computationally heavy. Instead, we used the \textit{alternative search} \citep{she2009thresholding,she2010}, which has two steps, choosing $\lambda_2$ while fixing $\lambda_1$ to a small constant and choosing $\lambda_1$ with the choice of $\lambda_2$. In the first step, due to a small magnitude of $\lambda_1$, the SGL perform like a group lasso, i.e., performing group selection but not much the within-group selection, which is somewhat comparable to the group selection step in the $gOMP$-$Thresholding$ (that corresponds Line 4 of Algorithm 1). Figure \ref{Reconstruct_Compare} compares the numerical outcome of the SGL to that of the $gOMP$-$Thresholding$ with no thresholding steps. Both are comparable to each other. 

\begin{figure}
	\centering
	\begin{subfigure}{0.6\textwidth}
		\includegraphics[width=\textwidth]{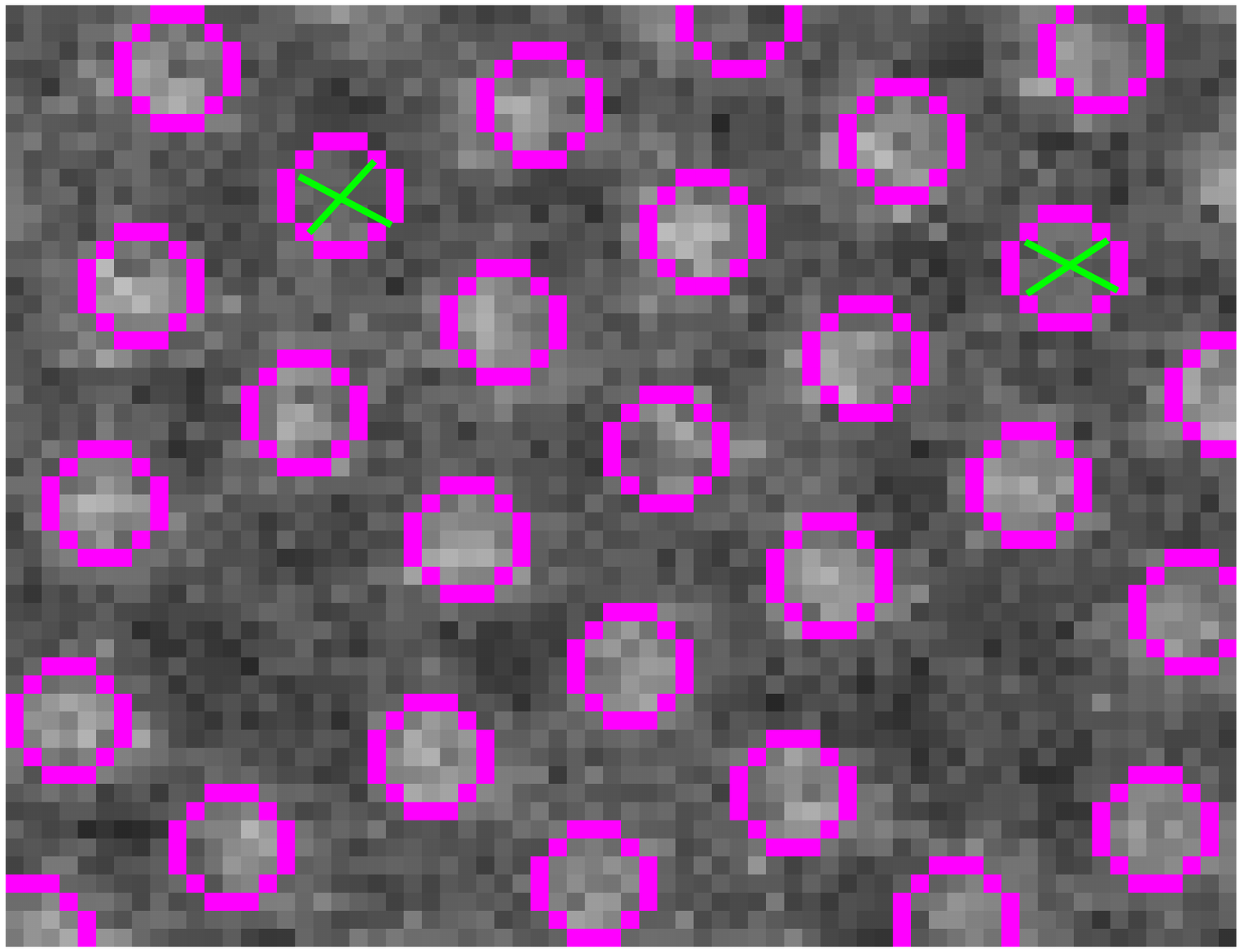}
		\caption{}
	\end{subfigure}
	\begin{subfigure}{\textwidth}
		\includegraphics[width=\textwidth]{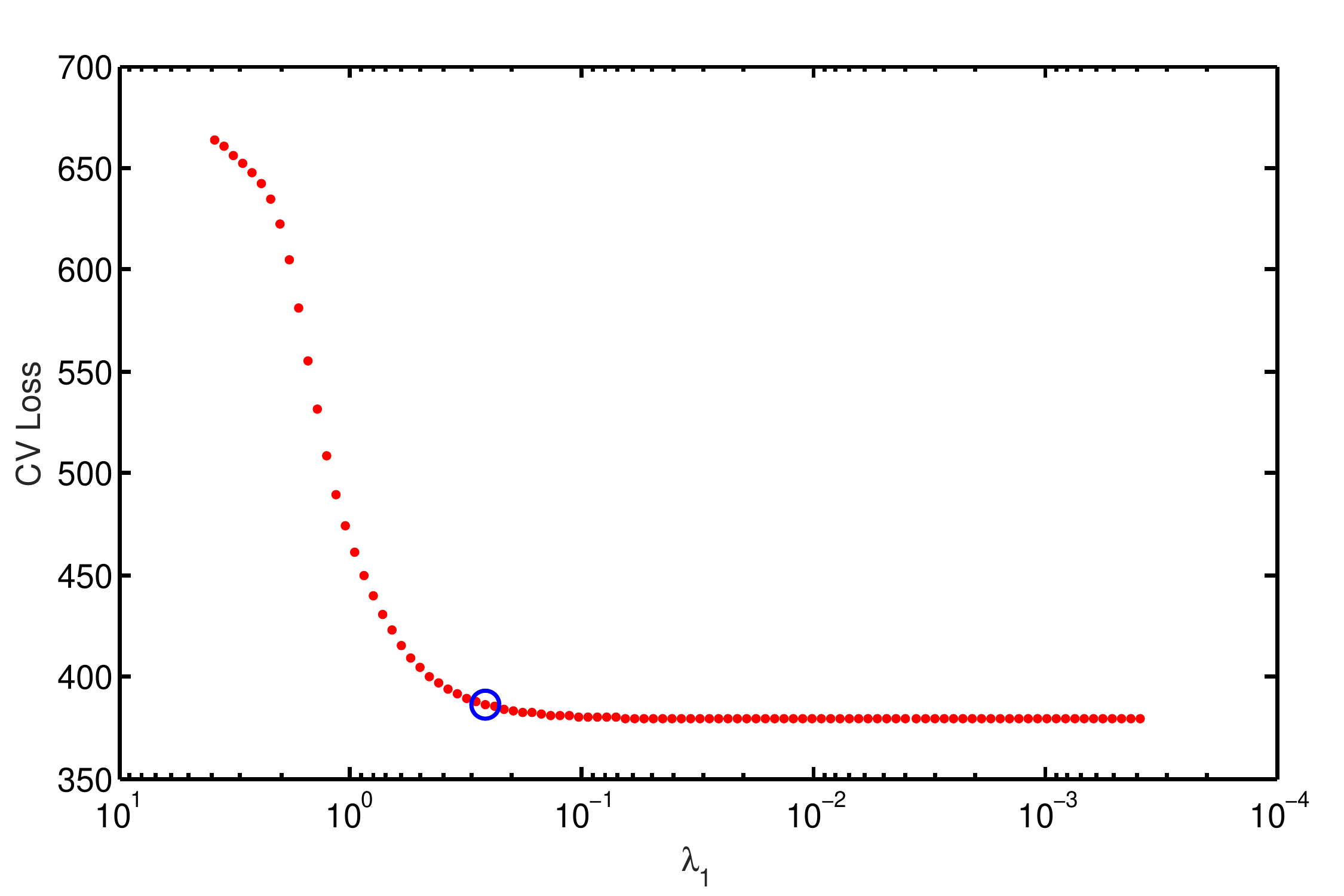}
		\caption{}
	\end{subfigure}
	\caption{results of the sparse group lasso (SGL) with $\lambda_1$ and $\lambda_2$ fine-tuned. (a) individual atom locations identified are marked with circles; false positives marked with crosses. (b) SCV loss versus $\lambda_1$; a circle locates the minimum SCV loss.}
	\label{results_Compare_SGL}
\end{figure}

\begin{figure}
	\centering
	\begin{subfigure}{0.6\textwidth}
		\includegraphics[width=\textwidth]{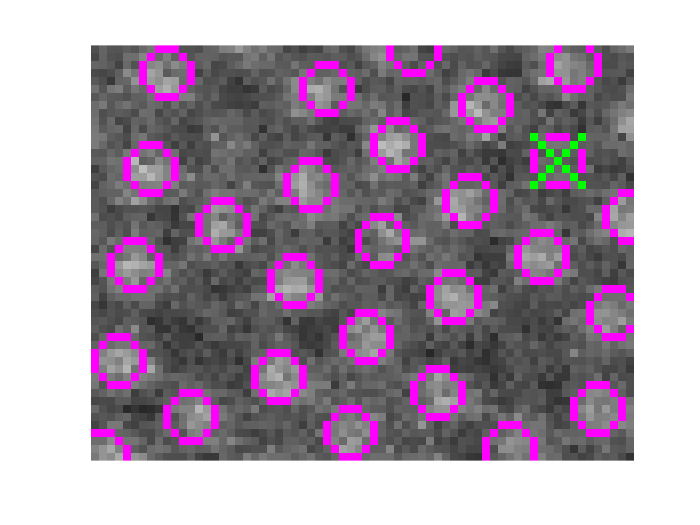}
		\caption{}
	\end{subfigure}
	\begin{subfigure}{\textwidth}
		\includegraphics[width=\textwidth]{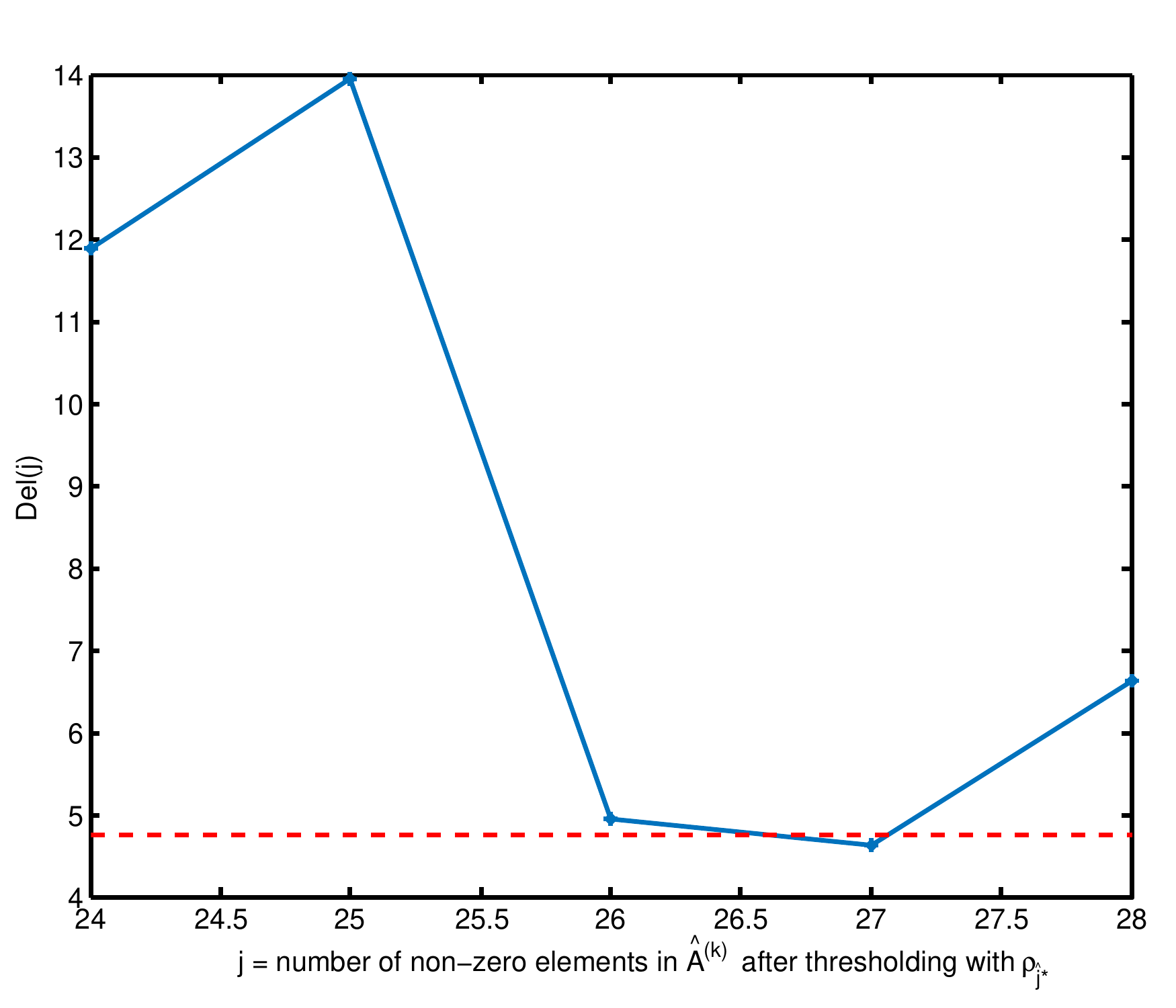}
		\caption{}
	\end{subfigure}
	\caption{results of the proposed approach with threshold $\rho$ chosen to $\rho_{\hat j^*}$. (a) individual atom locations identified are marked with circles, and false positives are marked with crosses. (b) $Del(j)$ versus $\hat\sigma\delta_k$ (horizontal bar) is plotted for threshold selection. $\hat{j}^*$ is selected to the first $j$ that achieves $Del(j)$ below the horizontal bar, following \eqref{rho_tuning}. }
	\label{results_Compare}
\end{figure}

Once $\lambda_2$ is chosen, $\lambda_1$ is fine-tuned using the selective cross validation \citep{she2013group}. We compared the SGL outcome with the choice (Figure \ref{results_Compare_SGL}) to the proposed $gOMP$-$Thresholding$ algorithm (Figure \ref{results_Compare}). The SGL made two false detections, while the proposed approach made only one false detection. We observed from many other numerical cases that choosing a good $\lambda_2$ for the SGL was not straightforward, while the proposed $gOMP$-$Thresholding$ algorithm has a good threshold selector as presented in Section \ref{sec:threshold}.

\section{Simulation study} \label{sec:sim}
\begin{figure}
	\centering
	\includegraphics[width=0.7\textwidth]{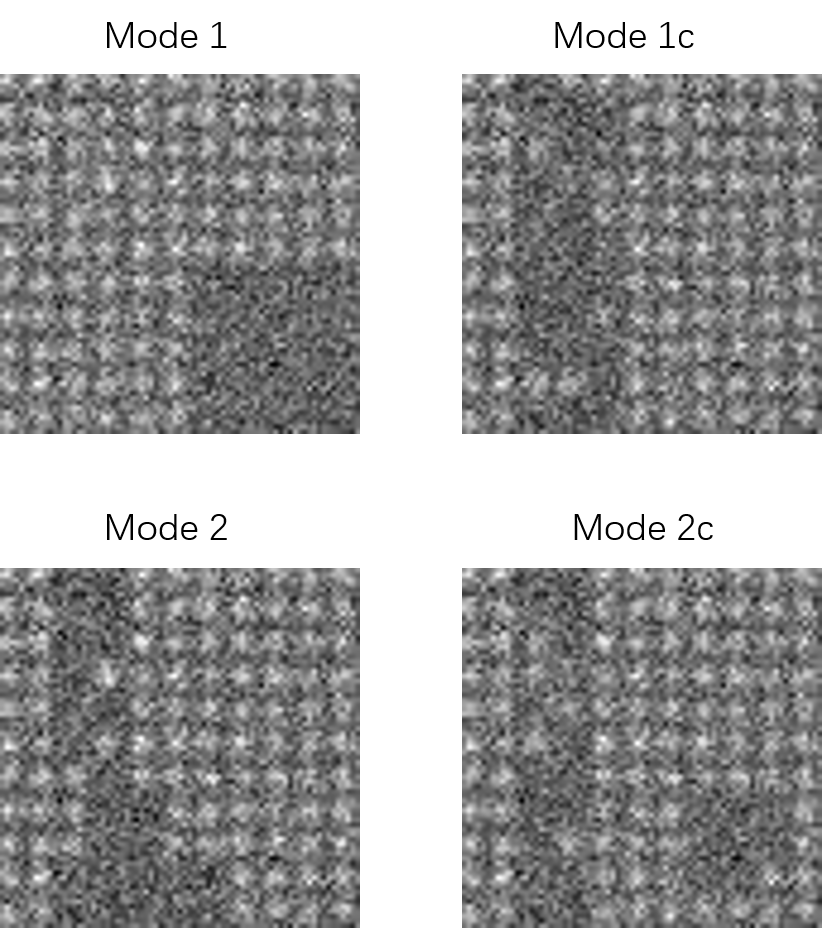}
	\caption{different spatial patterns of atom vacancy locations.}
	\label{modes}
\end{figure}

\begin{figure}
	\centering
	\includegraphics[width=\textwidth]{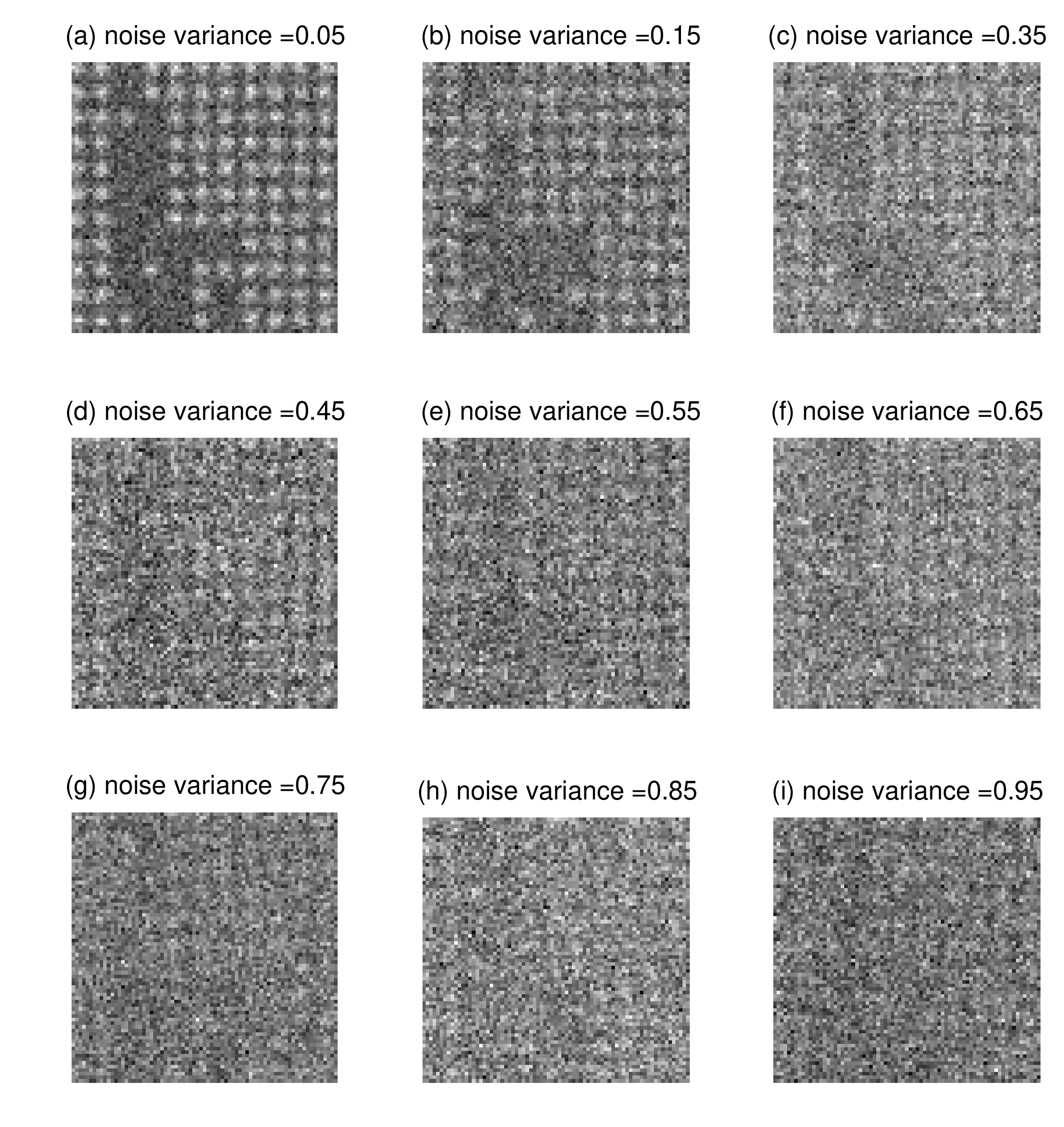}
	\caption{demonstration of noisy synthetic images used in our simulation study.}
	\label{noises}
\end{figure}

To understand how our method performs numerically, we performed intensive numerical experiments with synthetic images. A synthetic image sizes $75 \times 75$ in pixel, which can be occupied with 121 atoms if atoms locate on all lattice grid locations with no vacancy. We generated 100 random variations of the synthetic image. We considered three factors to generate the random variations. The first factor is the number of atom vacancies placed, which varied over $\{5, 10, 15, 20, 25\}$. The second factor is the spatial pattern of  atom vacancies. We considered five different patterns:  for the uniform mode, atom vacancies were uniformly sampled among 121 atom sites, and for the other four modes, they were randomly sampled among a subset of the 121 atom sites. Figure \ref{modes} shows the subsets for the four modes. For mode 1, potential vacancies are clustered in the bottom right of an image space, and atom vacancies are sub-sampled from those locations. The third design factor for the simulation is an observation noise level. We applied the Gaussian white noises with the variance in $\{0.05, 0.10, 0.15, 0.25, 0.35, 0.45, 0.55, 0.65, 0.75, 0.85, 0.95\}$. In this simulation study, the intensity scales of all synthetic images were normalized to $[0, 1]$, for which the true signal variance $\sigma_{sig}^2$ is around 0.075. The true signal variance was computed using synthetic images before noises are added. Since the signal-to-noise ratio for a noise variance $\sigma_{noise}^2$ can be calculated by $10\log_{10}(\sigma_{sig}^2) - 10 \log_{10}(\sigma_{noise}^2)$, the SNR values for the four noise variances we used are 1.7609, -3.0103, -5.2288, -6.6901, -7.7815,  -8.6530, -9.3785,-10.0000,-10.5436 and -11.0266 decibels respectively. The synthetic images we generated have extremely low contrasts as illustrated in Figure \ref{noises}. Combining different choices of the three factors generated 250 different simulation designs. 

For each design, we ran 50 replicated experiments, and the false positives and false negatives of atom detections were counted and averaged over the replicated experiments. The average counts were little influenced by the second factor (spatial patterns of atom vacancies). The variance of the average counts due to each factor was 137.4668 for the first factor (noise variance), 7.35 for the second factor (spatial pattern of vacancies), and 930.91 for the third factor (number of vacancies). The variance due to the second factor was quite comparable to the random variation of the average counts over 50 replicated experiments.

\begin{figure}
	\centering
	\includegraphics[width=\textwidth]{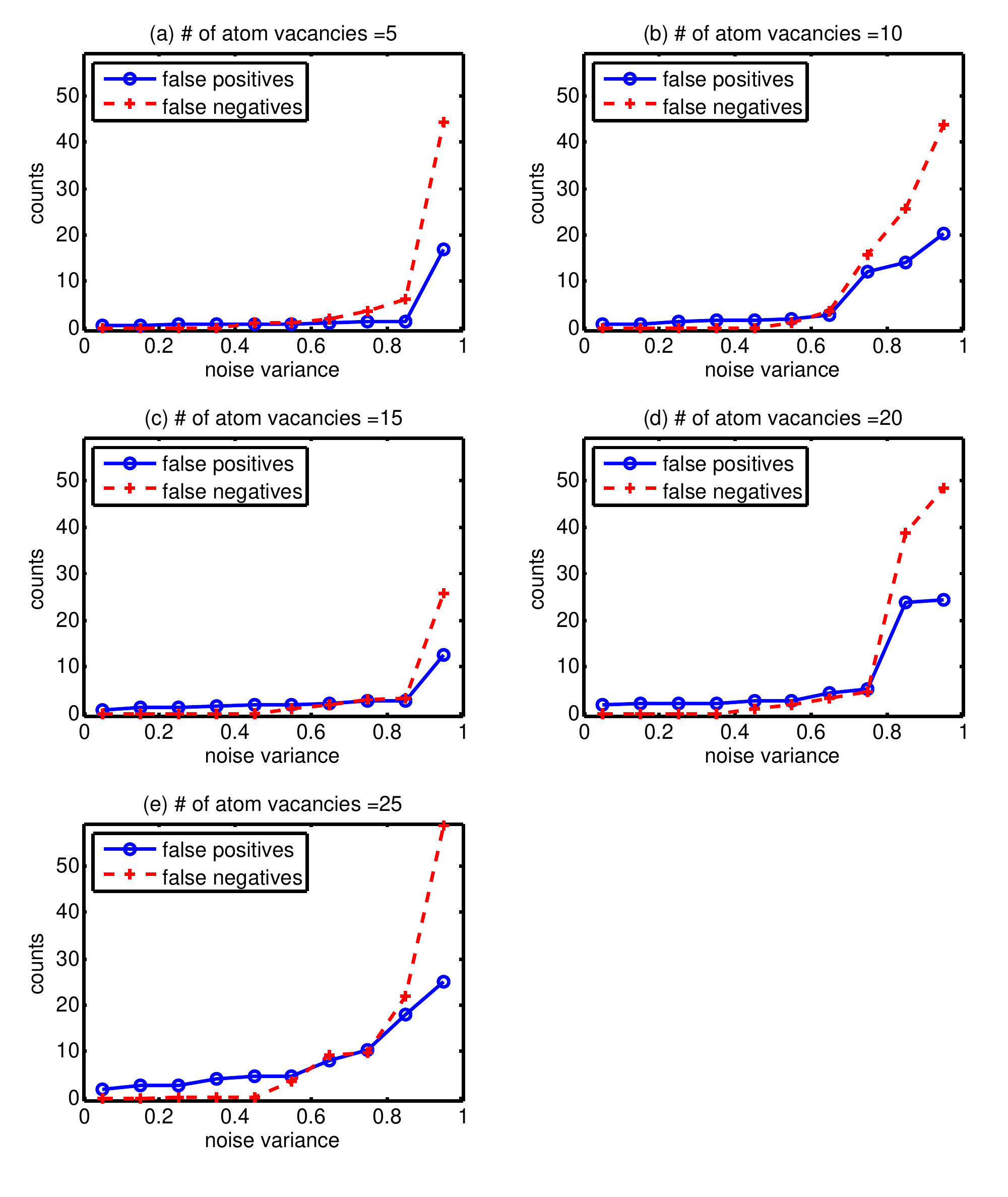}
	\caption{false positive and false negatives of estimated atom locations for different simulation designs}
	\label{FD_dis}
\end{figure}

\begin{figure}
	\centering
	\includegraphics[width=\textwidth]{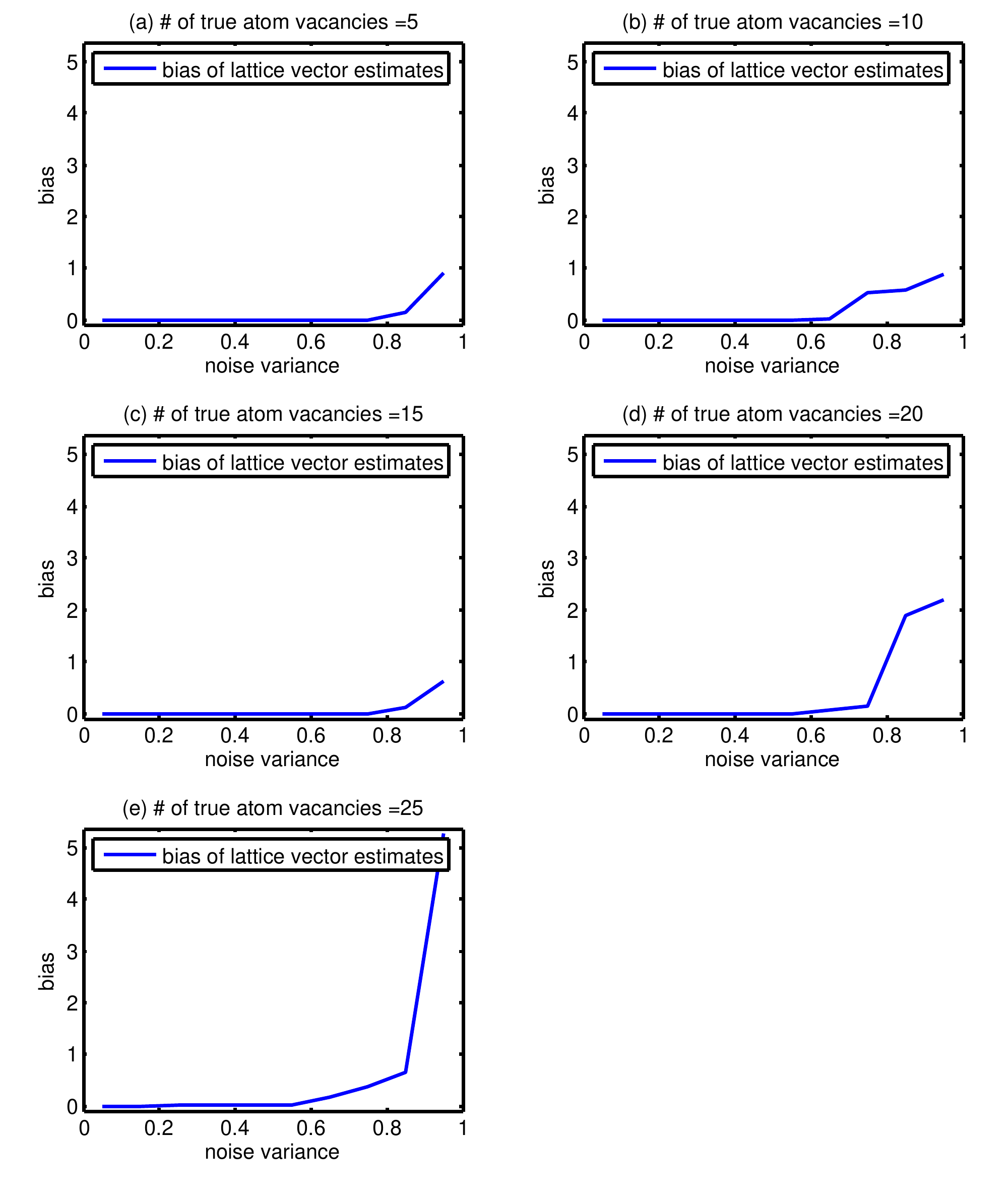}
	\caption{bias of the estimates of lattice basis $\V{p}_g$ and $\V{q}_g$ for different simulation designs. The bias presented here is defined as the total sum of individual biases of estimated $\V{p}_g$ and $\V{q}_q$, where an individual bias is the L2 norm of the difference of the estimated and true basis vector.}
	\label{GRP_dis}
\end{figure}

Figure \ref{FD_dis} summarizes the average counts of the false positives and false negatives for different levels of the first and third factors. The numbers of the false positives and the false negatives were kept very low and steady when the noise variance is below 0.6. However, the numbers increased significantly for the noise variances above 0.6. The major reason of the increase is related to the failure of estimating two lattice basis vectors, $\V{p}_g$ and $\V{q}_g$, under high image noises. To support this, we calculated the L2 norm of the difference of the estimate and the true value for each of the two lattice basis vectors, and defined the sum of the L2 norms as a bias. Figure \ref{GRP_dis} presents the biases for different noise variances and different numbers of atom voids. The biases were zero for noise variances below 0.6 but increased for higher noise variances. The lowest noise variances that give positive values of the bias were varied depending on the number of atom voids. Typically, with more voids, the bias started to increase in lower noise variances. If the biases of the basis vector estimates increase, a group of potential atom locations restricted by the basis vectors become misaligned to true atom locations, which causes rise in false positives and false negatives. 

%Figure \ref{ROC} shows the ROC curve of the false positive rates and false negative rates for 100 simulation designs.

%\begin{figure}
%	\centering
%	\includegraphics[width=\textwidth]{ROC.pdf}
%	\caption{ROC Curve of False Positive Rates and True Positive Rates}
%	\label{ROC}
%\end{figure}

\section{Application to analysis of Mo-V-M-O catalysts} \label{sec:app}
In this section, we present three real examples to demonstrate how our proposed approach is applied to the atomic-level structural determination of Mo-V-M-O oxide materials. We synthesized different Mo-V-M-O catalysts at the Oak Ridge National Laboratory following the methods reported in \citet{he2015better}. The materials were imaged at sub angstrom spatial resolution using a high angle annular dark field scanning transmission electron microscope (HAADF-STEM) at the Oak Ridge National Laboratory. The images were labeled as \texttt{B5}, \texttt{H5} and \texttt{H10} respectively. The signal-to-noise ratios for the three images were estimated to 7.38, 8.74 and 0.91 decibels respectively. For the ratio estimation, we handpicked several image foreground areas where atoms locate and several background areas where no atom locates. We computed the variance of the image intensities in the foreground areas and the variance of the intensities in the background areas. Assuming independence of the foreground signals and background, we can estimate the true signal variance by subtracting the background variance from the foreground variance, while the background variance becomes the noise variance. The last image has a very low SNR value. 

\begin{figure}
	\centering
	\includegraphics[width=0.7\textwidth]{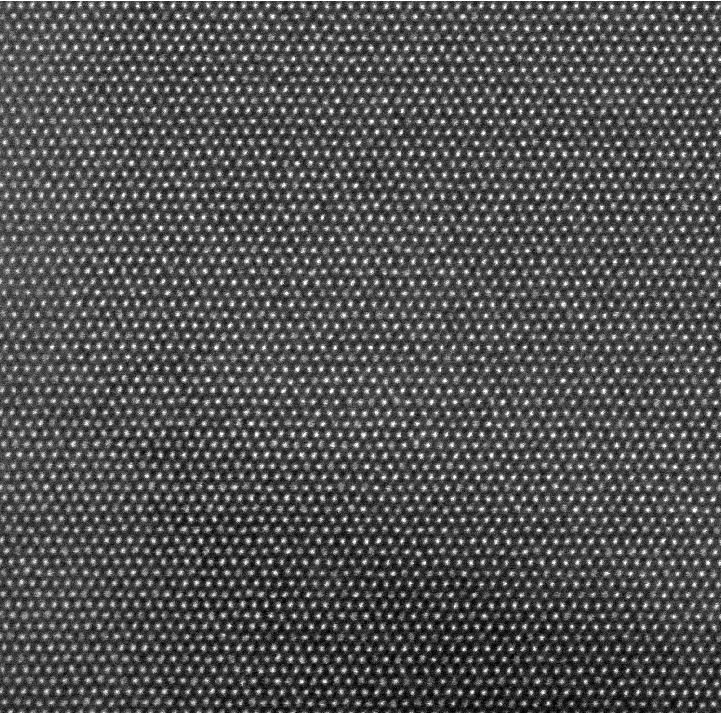}
	\caption{B5 image}
	\label{Raw-Grain}
\end{figure}

\begin{figure}
	\centering
	\includegraphics[width=0.7\textwidth]{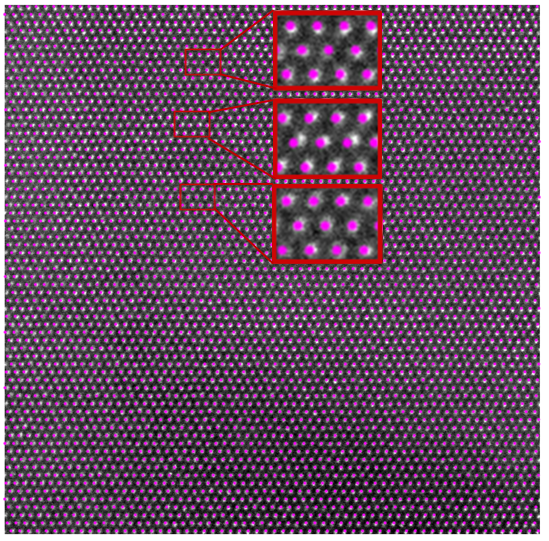}
	\caption{Estimated lattice on B5 image.}
	\label{Grain_results}
\end{figure}

The \texttt{B5} image in Figure \ref{Raw-Grain} was used to demonstrate the capability of our proposed approach to identify the global lattice grid of atoms in a sample Mo-V-M-O catalyst under local image distortions. Although a precision of a scanning electron microscope has been significantly advanced, the precise control of a electron probe at a sub-angstrom level is still challenging, so the probe location assigned for imaging may deviate from the actual probe location, which causes mild image distortions \citep{sang2016dynamic}. When such image distortions occurred during imaging a crystal material with single lattice grid, they could make the illusion that the sample material consists of multiple lattice grids, which would make a lattice identification difficult. With the group-level sparsity posed in our approach, the lattice identification can be very robust to local image distortions. Figure \ref{Grain_results} shows the outcome of our proposed approach with the \texttt{B5} image, where the estimated lattice positions were overlaid on the image. It can seen that most of the lattice positions identified (solid dots) match well to the actual atom locations (bright spots). A few exceptions are on the top left portion of the figure, where atoms are slightly off from the estimated lattice positions at a local area, and the deviations quickly disappear outside the local area. For a better presentation, we magnified some parts of the figure, which are shown in three boxes. In the middle box, atoms deviate from the lattice grid identified, but the deviations disappear or reduce in the upper and lower boxes. This is a local image distortion resulting from the scan ``ramp-up'' \citep{sang2016dynamic}, but detecting through simple visual inspection of the original image is very labor-intensive and subject to human errors. The proposed approach is capable of identifying a global lattice grid correctly under such local image distortions. This capability is practically meaningful since it can be used as a robust atomic lattice identification method. Moreover, the calculation of the deviations from a global lattice grid can provide important feedbacks to control an electron probe precisely. 
 
\begin{figure}
	\centering
	\begin{subfigure}{\textwidth}
		\includegraphics[width=0.9\textwidth]{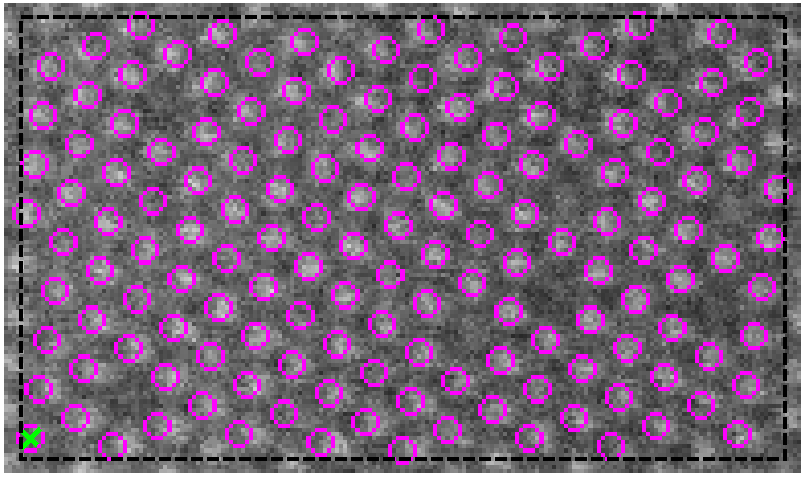}
		\caption{}
	\end{subfigure}
	\begin{subfigure}{\textwidth}
		\includegraphics[width=0.9\textwidth]{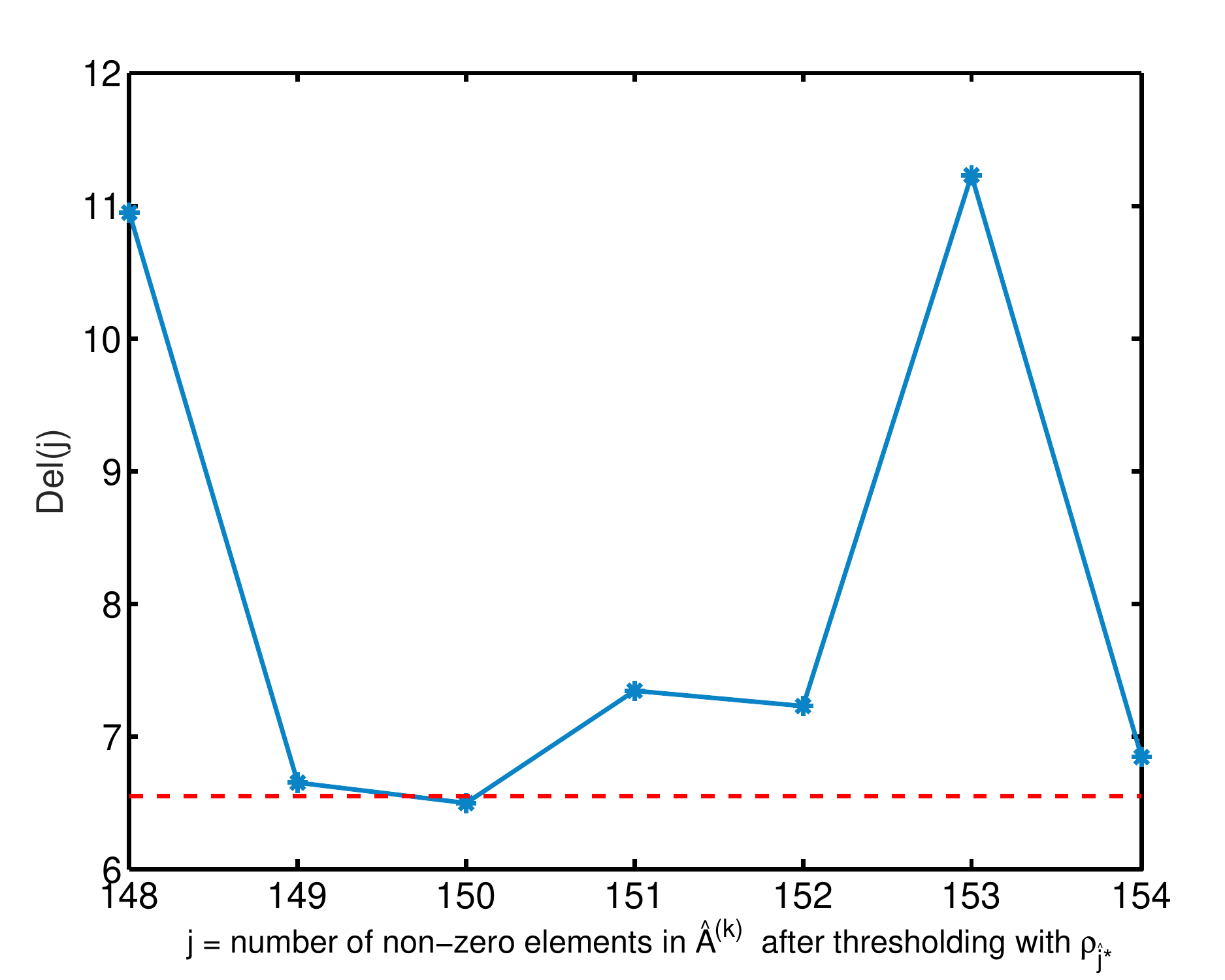}
		\caption{}
	\end{subfigure}
	\caption{Results of the proposed approach for \texttt{H5} with threshold $\rho$ chosen to $\rho_{\hat j^*}$; (a) individual atom locations identified are marked with circles while false positives are marked with crosses (to avoid any effects of atoms cropped around the image boundary, the image region outside the black dashed bounding box was not analyzed.) (b) $Del(j)$ versus $\hat\sigma\delta_k$ (horizontal bar) is plotted for threshold selection. $\hat{j}^*$ is selected to the first $j$ that achieves $Del(j)$ below the horizontal bar, following \eqref{rho_tuning}.}
	\label{H5}
\end{figure}

\begin{figure}
	\centering
	\begin{subfigure}{\textwidth}
		\includegraphics[width=0.9\textwidth]{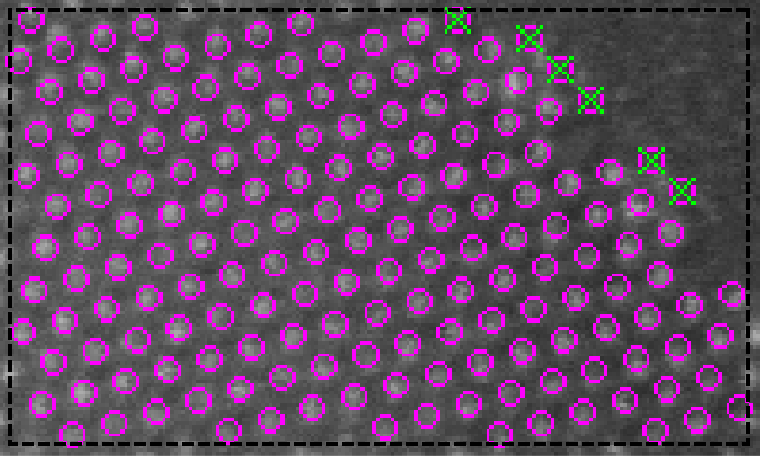}
		\caption{}
	\end{subfigure}
	\begin{subfigure}{\textwidth}
		\includegraphics[width=0.9\textwidth]{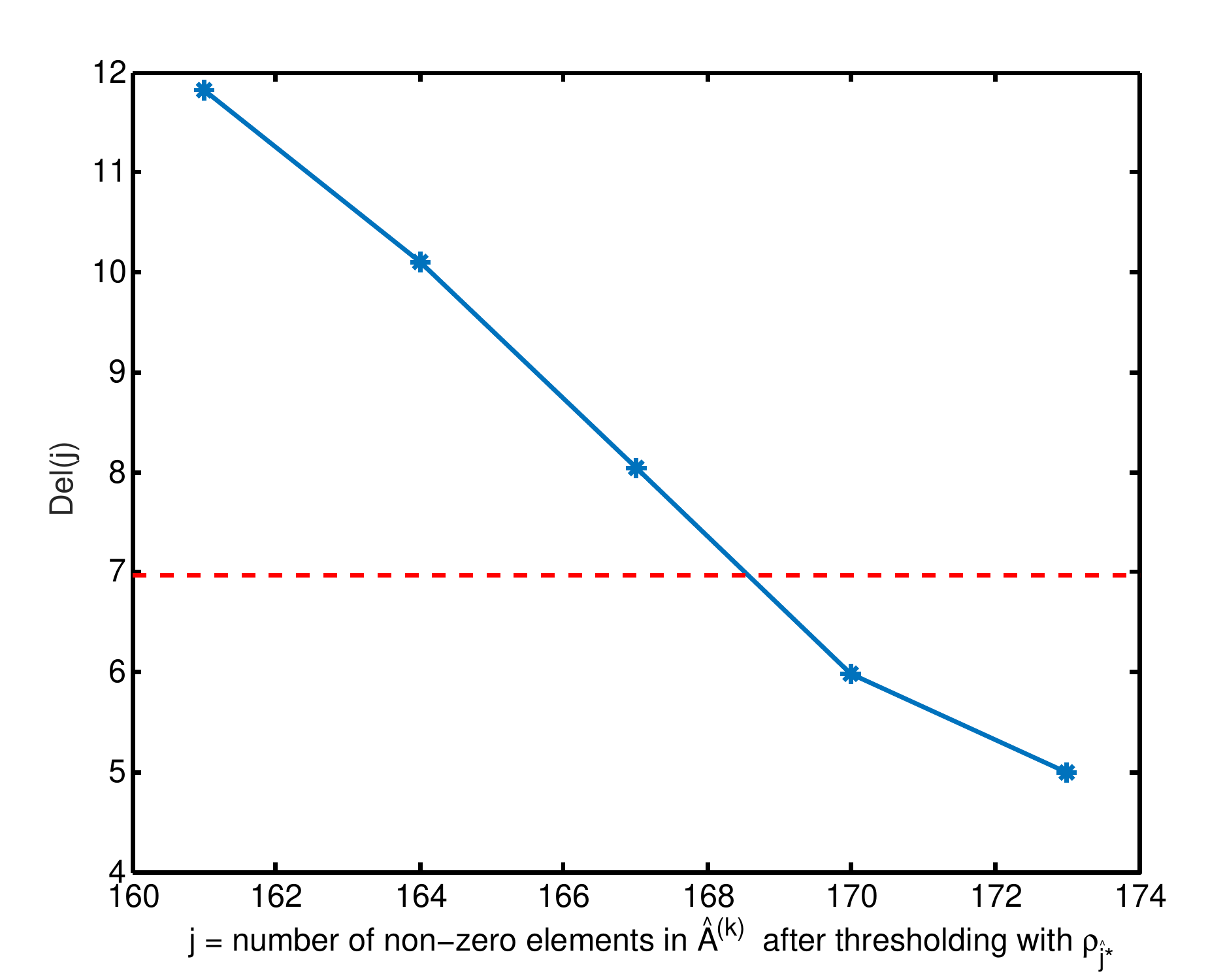}
		\caption{}
	\end{subfigure}
	\caption{results of the proposed approach for \texttt{H10} with threshold $\rho$ chosen to $\rho_{\hat j^*}$; (a) individual atom locations identified are marked with circles while false positives are marked with crosses (to avoid any effects of atoms cropped around the image boundary, the image region outside the black dashed bounding box was not analyzed.) (b) $Del(j)$ versus $\hat\sigma\delta_k$ (horizontal bar) is plotted for threshold selection. $\hat{j}^*$ is selected to the first $j$ that achieves $Del(j)$ below the horizontal bar, following \eqref{rho_tuning}.}
	\label{H10}
\end{figure}

Atom vacancies (regarded as defects in a crystal lattice) are another structure features closely related to material properties. For example, distribution of atom vacancies are also intrinsically coupled with magnetic, electronic and transport properties of solid-oxides \citep{adler2004factors}. We applied gOMP-thresholding algorithm to \texttt{H5} and \texttt{H10} images of Mo-V-M-O oxide materials for detecting atomic vacancies on the lattice.  Figure \ref{H5} shows the outcome of atom and vacancy detections for the \texttt{H5} image. The proposed algorithm identified an lattice group and detected atomic vacancies correctly. The number of the false negatives was zero, while there is one false positive. \texttt{H10} image has more complex patterns. Figure \ref{H10} shows the outcome of atom detections.  The half of image contains atoms and the other half is the background with no atoms. Our method still yielded impressive results with only six false positives out of 170 detections and did not produced any false negative. The sparse group lasso (SGL) produced more false positives and false negatives for the two test images, 93 false negatives and 0 false positives for \texttt{H5} and 8 false negatives and 2 false positives for \texttt{H10}. Again, we believe that the SGL result can be improved with a better choice of its tuning parameters. However, the parameter tuning is not straightforward, and the cross validation choice did not work very well. In summary, the gOMP-Threshold algorithm was successful for estimating atom vacancies in low-contrast STEM images, serving as a complementary tool with first principle approaches such as density function theory (DFT). A follow up work could be combining gOMP-thresholding and DFT to investigate atomic-defect configurations that are very important for future nano electronic devices and catalytic applications \citep{sang2016atomic}. 

The process of automatic atomic structure identification gets more important with the advent of genomic libraries, like the NIST Materials Genome Initiative \citep{dima2016} to determine the structure-property relations at the atomic scale. Results for H5 and H10 images indicate that our method can locate atoms and identify their spatial arrangements and existing defects in the arranging pattern. This allows us to incorporate our method with current work on nanophase material \citep{vasudevan2015big} to create a novel algorithm for lattice classification on multiple-phase material.

In terms of computation, the proposed approach took 18 minutes for analyzing the \texttt{H5} and \texttt{H10} images. The majority of the computing time was used to evaluate the threshold value described in expression \eqref{eq:thresh}. For the \texttt{B5} image, we have not included the thresholding, because we only need to the global lattice grid with no need for estimating atom vacancies. The computation time without thresholding was only 11 seconds. 

\section{Conclusion} \label{sec:conc}
The methodology proposed in this paper allows automated analysis of a atomically resolved images to locate individual atoms and identify their spatial symmetries and defects. Many atom locations can be succinctly represented by a lattice group spanned by a few basis vectors. Therefore, by identifying a few lattice groups exhibited in an input image, one can estimate most atom locations accurately. Identifying the underlying lattice groups among all possible groups was formulated as a sparse group selection problem. Posing only the sparse group regularization has a high risk of false positives. We further pose the within-group sparsity in addition to the group sparsity. The two-level sparsity was expressed as the regularization term in the form of the hierarchical tree-structured sparsity inducing norm. The two-level active set type algorithm was devised as a solution approach.  The proposed approach was validated with simulation and real datasets. 

To our best knowledge, this is among the first trials to analyze atomic scale images to extract the lattice structural information of materials. Feature database on structure information can be constructed based on our approach, which will lay out the foundation for structural study of materials from microscope image data. We believe this an exciting direction for developing quantitative methodologies to analyze atomically resolved  materials data, and expect these types of methodologies to become more popular in processing catalysts, 2D materials, complex oxides, and other varieties of engineering and scientifically relevant materials. 

\section*{Acknowledgments}
The authors are thankful for generous support of this work. Li and Park were partially supported by the National Science Foundation (NSF-1334012), and the Air Force Office of Scientific Research (AFOSR FA9550-13-1-0075 and AFOSR FA9550-16-1-0110). Belianinov, Dyck and Jesse were sponsored by the Laboratory Directed Research and Development Program of Oak Ridge National Laboratory, managed by UT-Battelle, LLC, for the U.S. Department of Energy. We gratefully acknowledge Albina Borisevich and Qian He from the Oak Ridge National Laboratory Materials Science and Technology Division for STEM micrographs of Mo-V-M and Mo-V-Te-Ta oxides.

\bibliographystyle{imsart-nameyear} % use style apalike.bst
\bibliography{lattice-structure} % find references in myrefs.bib
\end{document}